\documentclass[twocolumn,preprintnumbers,amsmath,amssymb,superscriptaddress]{revtex4-1}

\usepackage{amsmath,amsfonts,amssymb}
\usepackage{pdfpages}

\usepackage{graphicx}
\usepackage{dcolumn}
\usepackage{bm}

\makeatletter
\def\@seccntformat#1{}
\makeatother
\renewcommand{\numberline}[1]{}

\begin{document}

\title{Topological Insulators, Topological Crystalline Insulators, and Topological Kondo Insulators (Review Article)}

\author{M. Zahid Hasan} \affiliation{Joseph Henry Laboratories: Department of Physics, Princeton University, Princeton, NJ 08544, USA} \affiliation{Princeton Institute for the Science and Technology of Materials, School of Engineering and Applied Science, Princeton University, Princeton NJ 08544, USA}
\author{Su-Yang Xu} \affiliation{Joseph Henry Laboratories: Department of Physics, Princeton University, Princeton, NJ 08544, USA}
\author{Madhab Neupane} \affiliation{Joseph Henry Laboratories: Department of Physics, Princeton University, Princeton, NJ 08544, USA}

\begin{abstract}
\textbf{In this Book Chapter we briefly review the basic concepts defining topological insulators and elaborate on the key experimental results that revealed and established their symmetry protected (SPT) topological nature. We then present key experimental results that demonstrate magnetism, Kondo insulation, mirror chirality or topological crystalline order and superconductivity in spin-orbit topological insulator settings and how these new phases of matter arise through topological quantum phase transitions from Bloch band insulators via Dirac semimetals at the critical point.}
\end{abstract}

\maketitle

\tableofcontents

\section{Introduction}

\begin{figure}
\includegraphics[width=9cm]{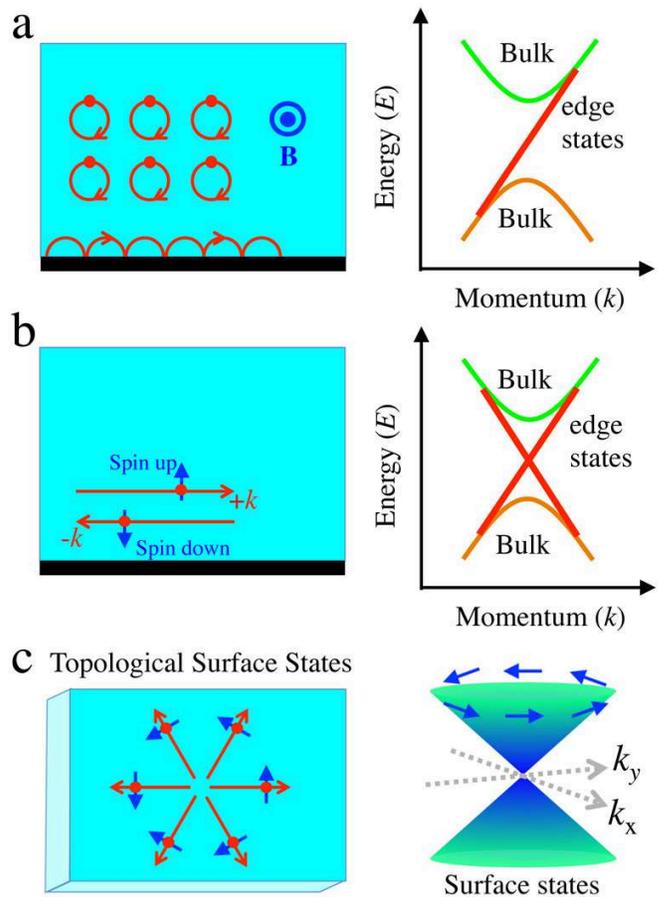}
\caption{\label{Cartoon1} \textbf{Topological insulators.} \textbf{a,} In the quantum Hall effect, the circular motion of electrons in a magnetic field, $\mathbf{B}$, is interrupted by the sample boundary. At the edge, electrons execute ``skipping orbits'' as shown, intimately leading to perfect conduction in one direction along the edge. \textbf{b,} The edge of the ``quantum spin Hall effect state'' or 2D topological insulator contains left-moving and right-moving modes that have opposite spin and are related by time-reversal symmetry. This edge can also be viewed as half of a quantum wire, which would have spin-up and spin-down electrons propagating in both directions. \textbf{c,} The surface of a 3D topological insulators supports electronic motion in any direction along the surface, but the direction of electron's motion uniquely determines its spin direction and vice versa. The 2D energy-momentum relation has a ``spin-Dirac cone'' structure but with helical or momentum space chiral spin texture with Berry's phase $\pi$ (spins go around in a closed loop in momentum space) \cite{RMP,Moore1}.}.
\end{figure}

Topological phases of matter differ from conventional materials in that a topological phase of matter features a nontrivial topological invariant in its bulk electronic wavefunction space which can be realized in a symmetry-protected condition \cite{RMP, Moore1,Zhang Physics Today, HasanMoore, Zhang_RMP, FuKM, Roy, 15, Hsieh1, Hsieh Science, TI book, TI book Zhang, Ando Review, Kimura Review}. The experimental discoveries of the 2D integer and fractional quantum Hall (IQH and FQH) states \cite{1,2,3,4,Tsui} in the 1980s realize the first two topological phases of matter in nature. These 2D topological systems are insulators in the bulk because the Fermi level is located in the middle of two Landau levels. On the other hand, the edges of these 2D topological insulators (IQH and FQH) feature chiral 1D metallic states, leading to remarkable quantized charge transport phenomena. The quantized transverse magneto-conductivity $\sigma_{xy} = ne^{2}/h$ (where $e$ is the electric charge and $h$ is Planck's constant) can be probed by charge transport experiments, which also provides a measure of the topological invariant (the Chern number) $n$ that characterizes these quantum Hall states \cite{5, 6}. In 2005, theoretical advances \cite{14,8} predicted a third type of 2D topological insulator, the quantum spin Hall (QSH) insulator. Such a topological state is symmetry protected. A QSH insulator can be viewed as two copies of quantum Hall systems that have magnetic field in the opposite direction. Therefore, no external magnetic field is required for the QSH phase, and the pair of quantum-Hall-like edge modes are related by the time-reversal symmetry (Fig.~\ref{Cartoon1}). The QSH phase was experimentally demonstrated in the mercury telluride quantum wells of using charge transport by measuring a longitudinal (charge) conductance of about $2e^2/h$ (two copies of quantum Hall states) at low temperatures \cite{7}. No spin polarization was measured in this experiment thus spin momentum locking which is essential for the Z$_2$ topological physics was not known or proven from experiments \cite{7}.

It is important to note that the 2D topological (IQH, FQH, and QSH) insulators are only realized at buried interfaces of ultraclean semiconductor heterostructures at very low temperatures \cite{7}. Furthermore, their metallic edge states can only be probed by the charge transport method \cite{7}. These facts hinder the systematic studies of many of their important properties, such as their electronic structure, spin polarization texture, tunneling properties, optical properties, as well as their responses under heterostructuring or interfacing with broken symmetry states. For example, the two counter-propagating edge modes in a QSH insulator is predicted to feature a 1D Dirac band crossing in energy and momentum space \cite{14, 8}. And edge mode moving along the $+k$ direction is expected to carry the opposite spin polarization as compared to that of moving to the $-k$ direction \cite{14, 8}. However, neither the Dirac band crossing nor the spin-momentum locking of the edge modes in a QSH insulator are experimentally observed, due to the lack of experimental probe that can measure these key properties for a 1D edge mode at a buried interface at mK temperatures, which is challenging. In 2007, it was also theoretically realized that the Z$_2$ topological number can have nontrivial generalization into three-dimensions \cite{FuKM, Fu:STI2,11,15}. In three-dimensions, there exist four (not three) Z$_2$ topological invariants that define the topological property of a 3D bulk material, namely $(\nu_0;\nu_1\nu_2\nu_3)$, where $\nu_0$ is the strong topological invariant, and $\nu_1- \nu_3$ are the weak topological invariants, respectively \cite{FuKM, Fu:STI2,11,15}. If the strong topological invariant is nonzero ($\nu_0=1$), the system is a 3D strong Z$_2$ topological insulator. It is important to note that the generalization from a 2D topological insulator (QSH) to a 3D strong Z$_2$ topological insulator is not a trivial generalization, because a 3D strong topological insulator cannot be adiabatically connected to multiple copies of 2D QSH insulators stacked along the out-of-plane $\hat{z}$ direction \cite{FuKM, Fu:STI2,11,15}. Therefore, the Z$_2$ topological order ($\nu_0=1$) in a 3D strong topological insulator represents a new type of genuinely three-dimensional (symmetry protected) topological order, which is fundamentally distinct from its 2D analogs (IQH, FQH, and QSH phases). The new topological order ($\nu_0=1$) leads to the existence of an odd number of gapless topological surfaces states at all surfaces of a strong topological insulator, irrespective of the choice of the surface termination \cite{FuKM, Fu:STI2,11,15, RMP}. These surface states are expected to be spin-momentum locked and their Fermi surfaces enclose the Kramers' points for an odd number of times \cite{RMP, FuKM, Hsieh1}. Moreover, they are protected by the time-reversal symmetry, which means that the topological surface states are robust against non-magnetic disorder and cannot be removed (gapped out) from the bulk band gap unless time-reversal symmetry is broken \cite{RMP, FuKM, Hsieh1}. Symmetry protected topological order is distinct from the fractional quantum Hall type topological order.

It turns out that the experimental discovery of the 3D topological insulator phase in 2007 \cite{Hsieh1} opened a new experimental era in fundamental topological physics. In contrast to its 2D analogs, (1) a 3D topological insulator can be realized at room temperatures without magnetic fields. Their metallic surface states exist at bare surfaces rather than only at buried interfaces \cite{RMP, Hsieh1}. (2) The electronic and spin groundstate of the topological surface states can be systematically studied by the spin- and angle-resolved photoemission spectroscopy (spin-ARPES) \cite{RMP, TI book, Hsieh1}, which provides a unique and powerful methodology for probing the topological order in three-dimensional topological phases. (3) Due to the relaxed conditions (room temperature, no magnetic field, bare surface), it is also possible to study the electrical transport, tunneling, optical, nanostructured, and many other key properties of the topological surface states \cite{RMP, TI book}. (4) The 3D topological insulator materials can be doped or interfaced to realize superconductivity or magnetism \cite{RMP, TI book}. (5) Since its discovery in 2007, there have been more than a hundred compounds identified as 3D topological insulators \cite{RMP, TI book}.

More importantly, the experimental discovery of 3D (Z$_2$) topological insulator \cite{10,Science, Xia, Zhang_nphys, Nature_2009, YLChen, Hsieh_PRL} has led to a surge of research in discovering other types of \textit{new} topological order in three-dimensions \cite{Xu, Ando QPT, TCI Story, TCI Hasan, TCI Ando, Vidya-1, CdAs_Hasan, CdAs_Cava, Chen_Na3Bi, Hasan N3Bi, Shi, MN, Feng, Golden}. The spin-resolved angle-resolved photoemission spectroscopy technique today constitutes a standard experimental methodology for discovering and probing new topological order (non-Z$_2$) in bulk solids \cite{Xu, Ando QPT, TCI Story, TCI Hasan, TCI Ando, Vidya-1, CdAs_Hasan, CdAs_Cava, Chen_Na3Bi, Hasan N3Bi, Shi, MN, Feng, Golden}. These fertile research frontiers include: (1) the topological crystalline insulator (TCI), where space group symmetries replace the time-reversal symmetry in a 3D Z$_2$ TI \cite{Liang PRL TCI, Liang NC SnTe}. The discovery of TCI \cite{TCI Story, TCI Hasan, TCI Ando, Vidya-1} following theoretical predictions \cite{Liang PRL TCI, Liang NC SnTe} leads to novel crystalline symmetry protected topological surface states. (2) the topological Kondo insulator (TKI), where the topological surface states in a TKI exist in the bulk Kondo gap rather than a simple Bloch gap in a Z$_2$ TI \cite{Dzero}. Demonstration of TKI \cite{Shi, MN, Feng, Golden} provides a platform for testing the interplay between topological order and strong electron correlation. (3) the topological Dirac/Weyl semimetals \cite{Wan, Nagaosa, Dirac_3D, Dirac_semi, Dai, CdAs_Hasan, CdAs_Cava, Chen_Na3Bi, Hasan N3Bi}, where new topological order (not $\nu_0$) can exist even if there is no global bulk energy gap, leading to multiple Dirac/Weyl nodes in the bulk and Fermi arc surface states on the surface \cite{Nagaosa}. (4) Superconducting \cite{Hor, Wray1, Morpurgo, BSCCO_Hasan, BSCCO_Valla} and magnetic \cite{Hor PRB BiMnTe, Wray2, Hedgehog, Checkelsky, Xue Science QAH} TIs and the topological phase transitions \cite{Xu, Ando QPT, Oh, Armitage}, which are the keys for a wide range of quantum phenomena such as Majorana fermion excitation \cite{Kane_Proximity}, topological magneto-electrical effect \cite{Qi PRB}, quantum anomalous Hall current \cite{Yu Science QAH}, as well as supersymmetry SUSY state \cite{SUSY}.

In this chapter, we review the experimental discoveries of symmetry-protected topologically (SPT) ordered phases in three-dimensions. We first review the discovery of 3D Z$_2$ topological insulator, which serves as the first topologically ordered phase of matter in 3D bulk materials. We elaborate the way of measuring the 3D Z$_2$ topological variant ($\nu_0=1$) by the spin- and angle-resolved photoemission spectroscopy (spin-ARPES) \cite{10, Science, Xia, Zhang_nphys, Nature_2009, YLChen, Hsieh_PRL}. In the following sections, we review the experimental efforts in discovering new topological order (non-Z$_2$) and new topological phenomena including Topological Kondo Insulators, Topological Quantum Phase Transition, Topological Dirac Semimetals, Magnetic and Superconductor Topological Insulators, and Topological Crystalline Insulators, respectively. The 3D topological materials are also experimentally studied by many groups world-wide using various techniques such as ARPES \cite{10,Science, Xia, Zhang_nphys, Nature_2009, YLChen, Hsieh_PRL, Xu, Ando QPT, TCI Story, TCI Hasan, TCI Ando, CdAs_Hasan, CdAs_Cava, Chen_Na3Bi, Hasan N3Bi, Shi, MN, Feng, Golden, Oh, Wray1, BSCCO_Hasan, BSCCO_Valla, Wray2, Hedgehog, Oh, Xue Nature physics QL, Ternary arXiv, Kuroda, Warping_Hasan, Warping_Ando, Chris, Preform, Hugo_Atomic, Gap, A Kimura, Valla, Ternary PRB, Preform, Gap, CD Gedik, Hofmann, Damascelli, CD_MN, 327, QL, Saddle, Rader}, scanning tunneling spectroscopies (STM) \cite{Roushan, Zhang_STM, Alpichshev, Cheng, Hanaguri, Haim Nature physics BiSe, Okada, Vidya-1, Vidya-2}, transport \cite{Qu, Analytis, Peng, Hadar1, Chen1, Ando_BTS, FuChun, Yayu, Fuhrer, Checkelsky, Fisk, Fisk2, Xue Science QAH, Samarth}, and optical methods \cite{Armitage, Hsieh_SHG, Hancock, Ultrafast, Ultrafast2, Light, Kerr, Basov}. Discovering and understanding topological ordered phases of matters in three-dimensions constitutes one of the most active research areas in condensed matter physics today.

\section{Z$_2$ topological insulators}

In this section, we review the experimental discovery of 3D Z$_2$ topological insulator, and elaborate the way of measuring the 3D Z$_2$ topological variant ($\nu_0=1$) by the spin- and angle-resolved photoemission spectroscopy (spin-ARPES). It was theoretically realized that strong spin-orbit coupling strength is one of the keys for realizing the 3D TI phase \cite{11,Murakami} since it leads to inversions between the bulk conduction and valence bands. The first 3D topological insulator is experimentally realized in the bismuth-antimony alloy system (Bi$_{1-x}$Sb$_x$). Bi$_{1-x}$Sb$_x$ is believed as a possible realization of 3D topological order for the following reason as predicted in band structure calculation \cite{11, Murakami, Lenoir, Liu, Wolff, Fukuyama, Buot}: Antimony (Sb) is a semimetal with strong spin-orbit interactions. Its bulk band electronic structure features one band inversion (an odd number) between the valence band maximum at the $T$ point of the bulk Brillouin zone (BZ). This fact makes antimony Z$_2$ topologically nontrivial ($\nu_0=1$) but antimony is a semimetal, which means there does not exist a full bulk band gap irrespective of the choice of the Fermi level. Substituting Sb by Bi is expected to change the relative energy levels of the bands at $T$ and $L$ points, and at antimony composition of $x\simeq0.1$, a full bulk energy gap is realized. Furthermore, it is also important to note that increasing Bi composition also effectively enhances the spin-orbit coupling. Thus for the system with very large bismuth composition ($0{\leq}x{\leq}0.04$), even the bands at the three $L$ points are inverted. Thus there are in-total four (an even number of) bulk band inversions in bismuth for the system with very large bismuth composition ($0{\leq}x{\leq}0.04$), making it (Z$_2$) topologically trivial. Therefore, theoretical band structure calculation predicts the 3D topological insulator phase in Bi$_{0.9}$Sb$_{0.1}$ ($x=0.1$).

\begin{figure*}
\includegraphics[width=17cm]{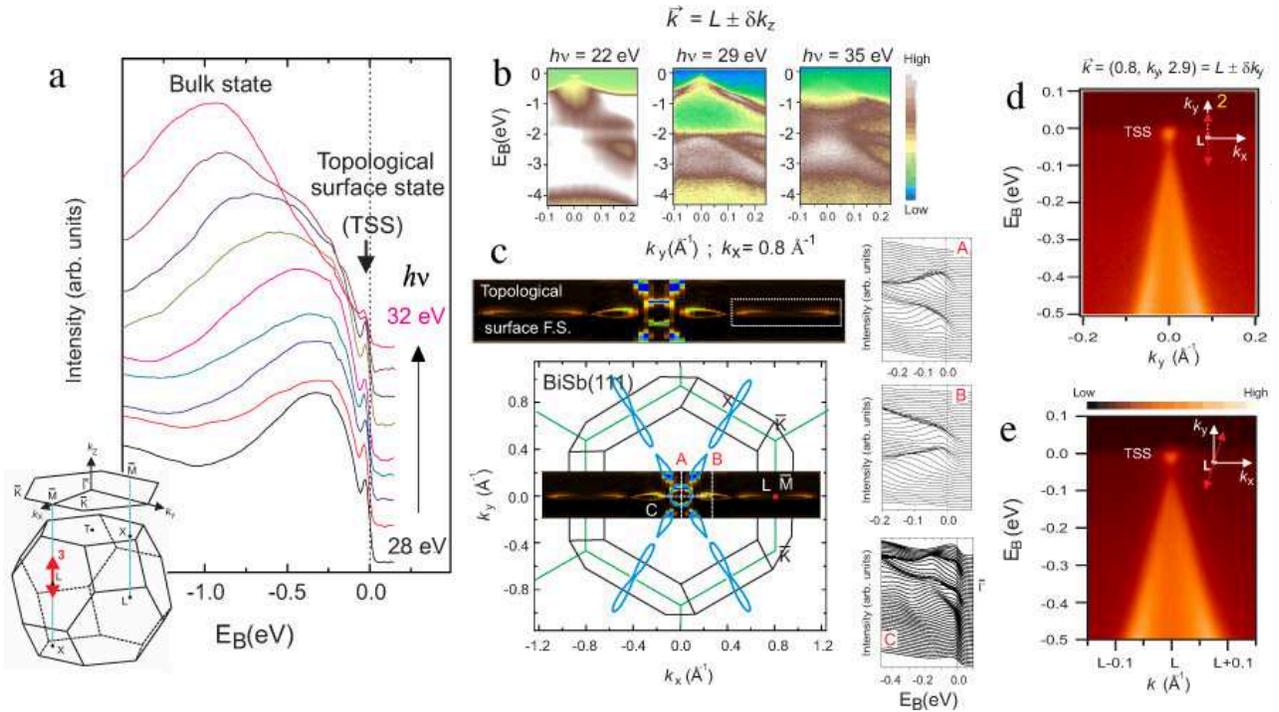}
\caption{\label{fig:BiSb_kz} \textbf{The first 3D topological insulator (2007): Topological Surface States and electronic band dispersion along the $\mathbf{k_z}$-direction in momentum space.} Surface states are experimentally identified by studying their out-of-plane momentum dispersion through the systematic variation of incident photon energy. \textbf{a,} Energy distribution curves (EDCs) of Bi$_{0.9}$Sb$_{0.1}$ with electrons at the Fermi level ($E_\textrm{F}$) maintained at a fixed in-plane momentum of ($k_x$=0.8 \AA$^{-1}$, $k_y$=0.0 \AA$^{-1}$) are obtained as a function of incident photon energy. \textbf{b,} ARPES intensity maps along cuts parallel to $k_y$ taken with electrons at $E_\textrm{F}$ fixed at $k_x$ = 0.8 \AA$^{-1}$ with respective photon energies of $h \nu$ = 22 eV, 29 eV and 35 eV. \textbf{c,} Projection of the bulk BZ (black lines) onto the (111) surface BZ (green lines). Overlay (enlarged in inset) shows the high resolution Fermi surface (FS) of the metallic SS mode, which was obtained by integrating the ARPES intensity (taken with $h \nu$ = 20 eV) from $-$15 meV to 10 meV relative to $E_\textrm{F}$. EDCs corresponding to the cuts A, B and C are also shown; these confirm the gapless character of the surface states in bulk insulating Bi$_{0.9}$Sb$_{0.1}$. \textbf{d,e,} ARPES dispersion cuts of Bi$_{0.9}$Sb$_{0.1}$. The cuts are along \textbf{d}, the $k_y$ direction, \textbf{e}, a direction rotated by approximately $10^{\circ}$ from the $k_y$ direction. [Adapted from D. Hsieh $et$ $al.$, \textit{Nature} \textbf{452}, 970 (2008) submitted in 2007 \cite{10}].}
\end{figure*}

In order to experimentally demonstrate the 3D topological insulator state in the Bi$_{0.9}$Sb$_{0.1}$ sample, we preform high-momentum-resolution angle-resolved photoemission spectroscopy (Fig.\ref{fig:BiSb_kz}) with varying incident photon energy (IPEM-ARPES). The incident photon energy dependent ARPES studies allow us to measure the energy dispersion along the out-of-plane momentum space direction ($E-k_{\perp}$), which can distinguish between the three-dimensional bulk bands and the two-dimensional surface states. As shown in Fig.\ref{fig:BiSb_kz} b, d, and e, a $\Lambda$-shaped dispersion whose tip lies less than 50 meV below the Fermi energy ($E_\textrm{F}$) is observed. Additional features originating from surface states that do not disperse with incident photon energy are also seen in Fig.\ref{fig:BiSb_kz} d and e. Our data are consistent with the extremely small effective mass of $0.002m_e$ (where $m_e$ is the electron mass) observed in magneto-reflection measurements on samples with $x = 11\%$ \cite{Hebel}. Studying the band dispersion perpendicular to the sample surface provides a way to differentiate bulk states from surface states in a 3D material. To visualize the near-$E_\textrm{F}$ dispersion along the 3D L-X cut (X is a point that is displaced from L by a $k_z$ distance of 3$\pi/c$, where $c$ is the lattice constant), in Fig.\ref{fig:BiSb_kz}a we plot energy distribution curves (EDCs), taken such that electrons at $E_\textrm{F}$ have fixed in-plane momentum $(k_x, k_y)$ = (L$_x$, L$_y$) = (0.8 \AA$^{-1}$, 0.0 \AA$^{-1}$), as a function of photon energy ($h\nu$). There are three prominent features in the EDCs: a non-dispersing, $k_z$ independent, peak centered just below $E_\textrm{F}$ at about $-$0.02 eV; a broad non-dispersing hump centered near $-$0.3 eV; and a strongly dispersing hump that coincides with the latter near $h\nu$ = 29 eV. To understand which bands these features originate from, we show ARPES intensity maps along an in-plane cut $\bar{K} \bar{M} \bar{K}$ (parallel to the $k_y$ direction) taken using $h\nu$ values of 22 eV, 29 eV and 35 eV, which correspond to approximate $k_z$ values of L$_z -$ 0.3 \AA$^{-1}$, L$_z$, and L$_z$ + 0.3 \AA$^{-1}$ respectively (Fig.\ref{fig:BiSb_kz}b). At $h\nu$ = 29 eV, the low energy ARPES spectral weight reveals a clear $\Lambda$-shaped band close to $E_\textrm{F}$. As the photon energy is either increased or decreased from 29 eV, this intensity shifts to higher binding energies as the spectral weight evolves from the $\Lambda$-shaped into a $\cup$-shaped band. Therefore, the dispersive peak in Fig.2a comes from the bulk valence band, and for $h\nu$ = 29 eV the high symmetry point L = (0.8, 0, 2.9) appears in the third bulk BZ. In the maps of Fig.\ref{fig:BiSb_kz}b with respective $h\nu$ values of 22 eV and 35 eV, overall weak features near $E_\textrm{F}$ that vary in intensity remain even as the bulk valence band moves far below $E_\textrm{F}$. The survival of these weak features over a large photon energy range (17 to 55 eV) supports their surface origin. The non-dispersing feature centered near $-0.3$ eV in Fig.\ref{fig:BiSb_kz}a comes from the higher binding energy (valence band) part of the full spectrum of surface states, and the weak non-dispersing peak at $-0.02$ eV reflects the low energy part of the surface states that cross $E_\textrm{F}$ away from the $\bar{M}$ point and forms the surface Fermi surface (Fig.\ref{fig:BiSb_kz}c).

We now discuss the topological character of surface states in Bi$_{0.9}$Sb$_{0.1}$ (Fig.\ref{fig:BiSb_kz}c), focusing on their key differences with respect to surface states in a conventional (topologically trivial) insulator. In general, surface states are allowed to exist within the bulk energy gap owing to the loss of space inversion symmetry $[E(k,\uparrow) = E(-k,\uparrow)]$. However, there is a key distinction between surface states in a conventional insulator and a topological insulator, which is that along a path connecting two TRIM in the same BZ, the Fermi energy inside the bulk gap will intersect these singly degenerate surface states either an even or odd number of times. If there are an even number of surface state crossings, the surface states are topologically Z$_2$ trivial because disorder or correlations can remove \emph{pairs} of such crossings by pushing the surface bands entirely above or below $E_\textrm{F}$. When there are an odd number of crossings, however, at least one surface state must remain gapless, which makes it non-trivial \cite{11, Murakami, Fu:STI2}. The existence of such topologically non-trivial surface states can be theoretically predicted on the basis of the \emph{bulk} band structure only, using the Z$_2$ invariant. Materials with band structures with Z$_2 = +1$  ($\nu_0$ = 0) are ordinary Bloch band insulators that are topologically equivalent to the filled shell atomic insulator, and are predicted to exhibit an even number (including zero) of surface state crossings. Materials with bulk band structures with Z$_2 = -1$ ($\nu_0$ = 1) on the other hand, which are expected to exist in rare systems with strong spin-orbit coupling acting as an internal magnetic field on the electron system \cite{Haldane(P-anomaly)}, and inverted bands at an odd number of high symmetry points in their bulk 3D BZs, are predicted to exhibit an odd number of surface state crossings, precluding their adiabatic continuation to the atomic insulator \cite{11, Murakami, Fu:STI2, 15, 8, 7}. Such topological surface states that enclose the Kramers' points by an odd number of times \cite{Fu:STI2, 15} cannot be realized in any purely 2D electron gas system, such as the one realized at the interface of GaAs/GaAlAs systems.

\begin{figure*}
\includegraphics[scale=0.56,clip=true, viewport=0.0in 0.5in 11.3in 6.5in]{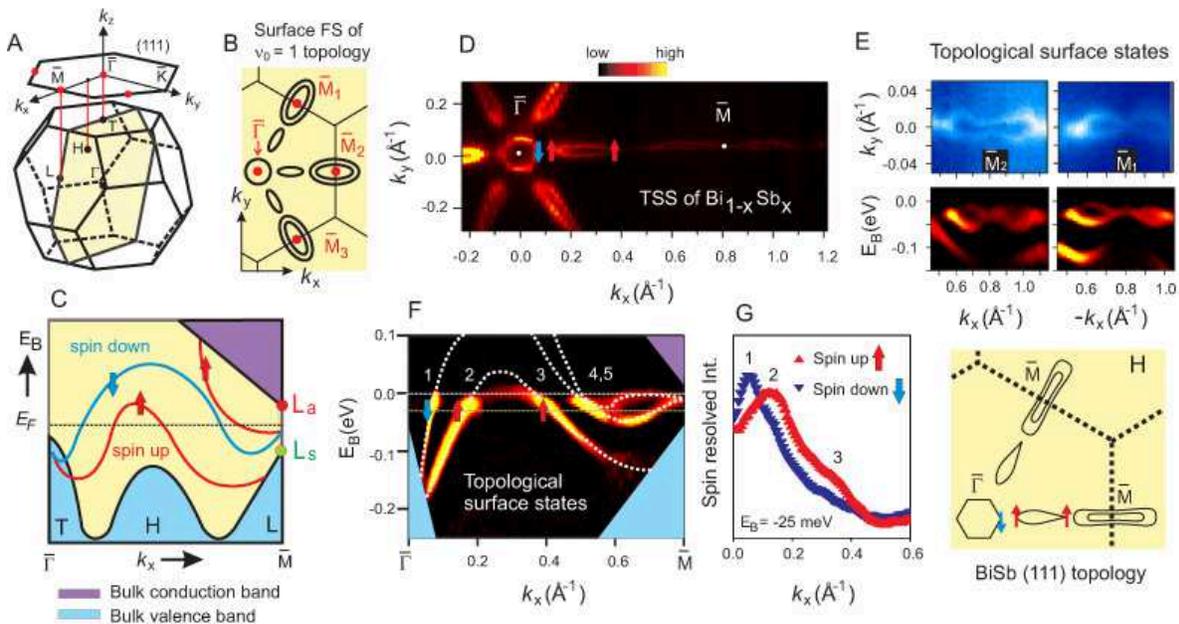}
\caption{\label{BiSb_spin} \textbf{Spin texture of a topological insulator encodes Z$_2$ topological order of the bulk (2008)} (A) Schematic sketches of the bulk Brillouin zone (BZ) and (111) surface BZ of the Bi$_{1-x}$Sb$_x$ crystal series.(B) Schematic of Fermi surface pockets formed by the surface states (SS) of a topological insulator that carries a Berry's phase. (C) Partner switching band structure topology. (D) Spin-integrated ARPES intensity map of the SS of Bi$_{0.91}$Sb$_{0.09}$ at $E_\textrm{F}$. Arrows point in the measured direction of the spin. (E) High-resolution ARPES intensity map of the SS at $E_\textrm{F}$ that enclose the $\bar{M}_1$ and $\bar{M}_2$ points. Corresponding band dispersion (second derivative images) are shown below. The left right asymmetry of the band dispersions are due to the slight offset of the alignment from the $\bar{\Gamma}$-$\bar{M}_1$($\bar{M}_2$) direction. (F) Surface band dispersion image along the $\bar{\Gamma}$-$\bar{M}$ direction showing five Fermi level crossings. (G) Spin-resolved momentum distribution curves presented at $E_\textrm{B}$ = $-$25 meV showing single spin degeneracy of bands at 1, 2 and 3. Spin up and down correspond to spin pointing along the +$\hat{y}$ and -$\hat{y}$ direction respectively. (H) Schematic of the spin-polarized surface FS observed in our experiments. It is consistent with a $\nu_0$ = 1 topology (compare (B) and (H)). [Adapted from D. Hsieh $et$ $al.$, \textit{Science} \textbf{323}, 919 (2009)\cite{Science}, Submitted in 2008].} \end{figure*}

The nontrivial Z$_2$ topological number ($\nu_0$ = 1) in a 3D topological insulator requires the terminated surface to have a Fermi surface (FS) that supports a non-zero Berry's phase (odd as opposed to even multiple of $\pi$), which is not realizable in an ordinary spin-orbit material. More specifically, for the Z$_2$ TI phase in Bi$_{1-x}$Sb$_x$, according to Kramers theorem, they must remain spin degenerate at four special time reversal invariant momenta ($\vec{k}_T$ = $\bar{\Gamma}$, $\bar{M}$) in the (111) surface BZ of Bi$_{1-x}$Sb$_x$ [see Fig.\ref{BiSb_spin}(A)]. If a Fermi surface pocket does not enclose $\vec{k}_T$ (= $\bar{\Gamma}$, $\bar{M}$), it is irrelevant for the Z$_2$ topology \cite{11,20}. Because the wave function of a single electron spin acquires a geometric phase factor of $\pi$ \cite{16} as it evolves by 360$^{\circ}$ in momentum space along a Fermi contour enclosing a $\vec{k}_T$, an odd number of Fermi pockets enclosing $\vec{k}_T$ in total implies a $\pi$ geometrical (Berry's) phase \cite{11}. In order to realize a $\pi$ Berry's phase the surface bands must be spin-polarized and exhibit a partner switching \cite{11} dispersion behavior between a pair of $\vec{k}_T$. This means that any pair of spin-polarized surface bands that are degenerate at $\bar{\Gamma}$ must not re-connect at $\bar{M}$, or must separately connect to the bulk valence and conduction band in between $\bar{\Gamma}$ and $\bar{M}$. The partner switching behavior is realized in Fig. \ref{BiSb_spin}(C) because the spin down band connects to and is degenerate with different spin up bands at $\bar{\Gamma}$ and $\bar{M}$.

\begin{figure}
\includegraphics[width=9cm]{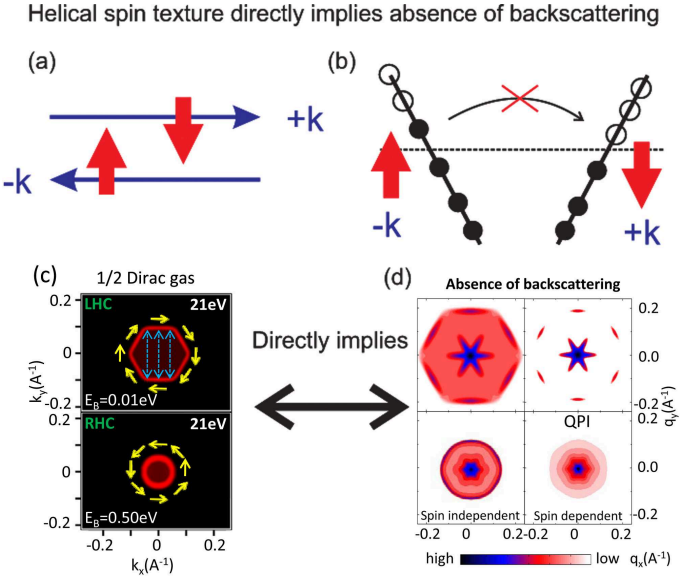}
\caption{\label{backscatter} \textbf{Helical spin texture naturally leads to absence of elastic backscattering for surface transport: No ``U'' turn on a 3D topological insulator surface.} (a) Our measurement of a helical spin texture in both Bi$_{1-x}$Sb$_x$ and in Bi$_2$Se$_3$ directly shows that there is (b) an absence of backscattering. (c) ARPES measured FSs are shown with spin
directions based on polarization measurements. L(R)HC stands for left(right)-handed chirality. (d) Spin independent and spin dependent scattering profiles on FSs in (c) relevant for surface quasi-particle transport are shown which is sampled by the quasi-particle interference (QPI) modes. [Adapted from S.-Y. Xu $et$ $al.$, \textit{Science} \textbf{332} 560 (2011). \cite{Xu}}
\end{figure}

We, for the first time, investigated the spin properties of the topological insulator phase \cite{Science}, in order to experimentally demonstrate the non-zero Berry's phase and the nontrivial Z$_2$ topological invariant. Spin-integrated ARPES \cite{19} intensity maps of the (111) surface states of insulating Bi$_{1-x}$Sb$_x$ taken at the Fermi level ($E_\textrm{F}$) [Figs \ref{BiSb_spin}(D)\&(E)] show that a hexagonal FS encloses $\bar{\Gamma}$, while dumbbell shaped FS pockets that are much weaker in intensity enclose $\bar{M}$. By examining the surface band dispersion below the Fermi level [Fig.\ref{BiSb_spin}(F)] it is clear that the central hexagonal FS is formed by a single band (Fermi crossing 1) whereas the dumbbell shaped FSs are formed by the merger of two bands (Fermi crossings 4 and 5) \cite{10}. This band dispersion resembles the partner switching dispersion behavior characteristic of topological insulators. To check this scenario and determine the topological index $\nu_0$, we have carried out spin-resolved photoemission spectroscopy. Fig.\ref{BiSb_spin}(G) shows a spin-resolved momentum distribution curve taken along the $\bar{\Gamma}$-$\bar{M}$ direction at a binding energy $E_\textrm{B}$ = $-$25 meV [Fig.\ref{BiSb_spin}(G)]. The data reveal a clear difference between the spin-up and spin-down intensities of bands 1, 2 and 3, and show that bands 1 and 2 have opposite spin whereas bands 2 and 3 have the same spin (detailed analysis discussed later in text). The former observation confirms that bands 1 and 2 form a spin-orbit split pair, and the latter observation suggests that bands 2 and 3 (as opposed to bands 1 and 3) are connected above the Fermi level and form one band. This is further confirmed by directly imaging the bands through raising the chemical potential via doping. Irrelevance of bands 2 and 3 to the topology is consistent with the fact that the Fermi surface pocket they form does not enclose any $\vec{k}_T$. Because of a dramatic intrinsic weakening of signal intensity near crossings 4 and 5, and the small energy and momentum splitting of bands 4 and 5 lying at the resolution limit of modern spin-resolved ARPES spectrometers, no conclusive spin information about these two bands can be drawn from the methods employed in obtaining the data sets in Figs \ref{BiSb_spin}(G)\&(H). However, whether bands 4 and 5 are both singly or doubly degenerate does not change the fact that an odd number of spin-polarized FSs enclose the $\vec{k}_T$, which provides evidence that Bi$_{1-x}$Sb$_x$ has $\nu_0$ = 1 and that its surface supports a non-trivial Berry's phase. This directly implies an absence of backscattering in electronic transport along the surface (Fig.\ref{backscatter}), which has been re-confirmed by numerous scanning tunneling microscopy studies that show quasi-particle interference patterns that can only be modeled assuming an absence of backscattering \cite{Roushan,Alpichshev,Zhang_STM}. More importantly, the spin-ARPES method that we developed in Ref. \cite{Science} becomes a standard experimental methodology for discovering and probing topological order (non-Z$_2$) in bulk solids \cite{TCI Hasan, CdAs_Hasan, MN}.

\begin{figure*}
\includegraphics[width=17cm]{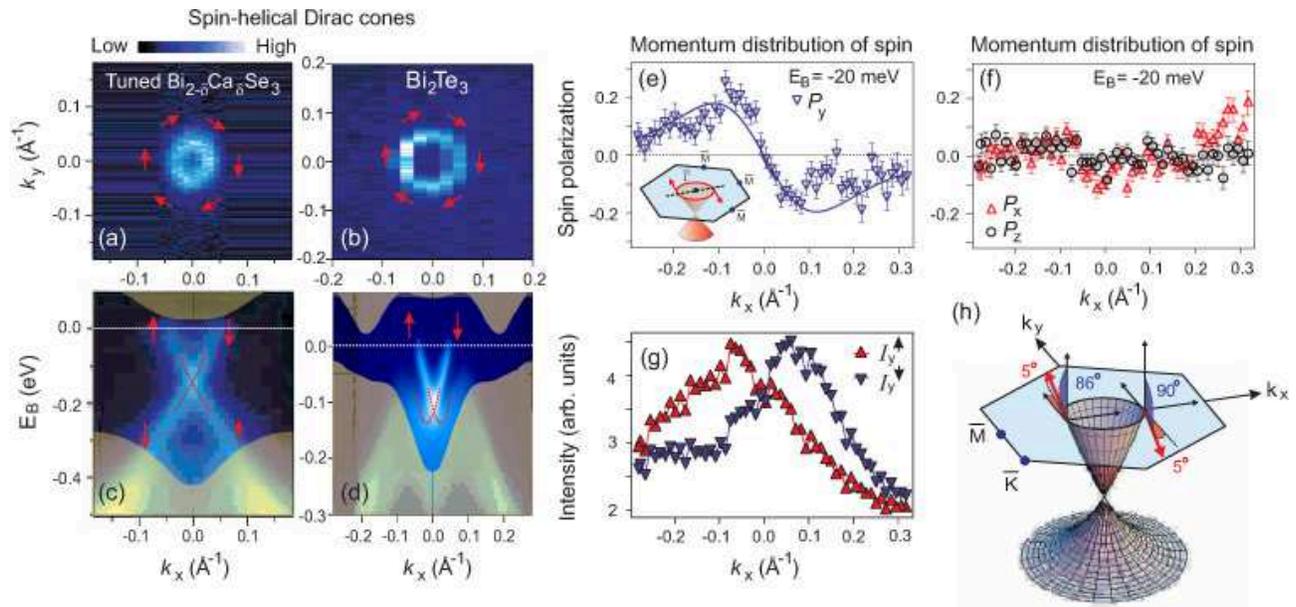}
\caption{\label{BiTe_spin} \textbf{First detection of Z$_2$ (symmetry protected) Topological-Order: spin-momentum locking of spin-helical Dirac electrons in Bi$_2$Se$_3$ and Bi$_2$Te$_3$ using spin-resolved ARPES.} (a) ARPES intensity map at $E_\textrm{F}$ of the (111) surface of tuned Bi$_{2-\delta}$Ca$_{\delta}$Se$_3$ (see text) and (b) the (111) surface of Bi$_2$Te$_3$. Red arrows denote the direction of spin around the Fermi surface. (c) ARPES dispersion of tuned Bi$_{2-\delta}$Ca$_{\delta}$Se$_3$ and (d) Bi$_2$Te$_3$ along the $k_x$ cut. The dotted red lines are guides to the eye. (e) Measured $y$ component of spin-polarization along the $\bar{\Gamma}$-$\bar{M}$ direction at $E_\textrm{B}$ = -20 meV, which only cuts through the surface states. Inset shows a schematic of the cut direction. (f) Measured $x$ (red triangles) and $z$ (black circles) components of spin polarization along the $\bar{\Gamma}$-$\bar{M}$ direction at $E_\textrm{B}$ = -20 meV. (g) Spin-resolved spectra obtained from the $y$ component spin polarization data. (h) Fitted values of the spin polarization vector \textbf{P}. [Adapted from D. Hsieh $et$ $al.$, \textit{Nature} \textbf{460}, 1101 (2009). Some data are adapted from Y. Xia $et$ $al.$, \textit{Nature Physics} \textbf{5}, 398 (2009)].}
\end{figure*}

\begin{figure*}
\includegraphics[width=17cm]{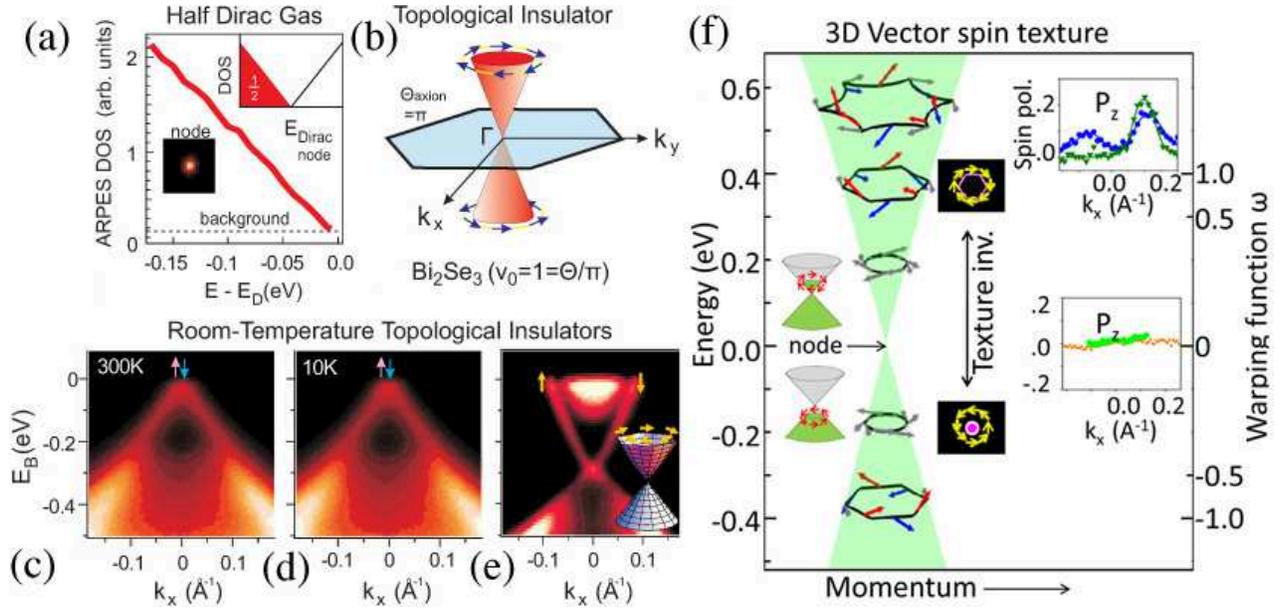}
\caption{\label{RT} \textbf{Observation of room temperature (300K) topological order (symmetry protected) without applied magnetic field in Bi$_2$Se$_3$:} (a) Crystal momentum integrated ARPES data near Fermi level exhibit linear fall-off of density of states, which, combined with the spin-resolved nature of the states, suggest that a half Fermi gas is realized on the topological surfaces. (b) Spin texture map based on spin-ARPES data suggest that the spin-chirality changes sign across the Dirac point. (c) The Dirac node remains well defined up a temperature of 300 K suggesting the stability of topological effects up to the room temperature. (d) The Dirac cone measured at a temperature of 10 K. (e) Full Dirac cone. (f) The spin polarization momentum-space texture as a function of energy with respect to the Dirac point. [Adapted from D. Hsieh $et$ $al.$, \textit{Nature} \textbf{460}, 1101 (2009). \cite{Nature_2009}].} \end{figure*}

It is worth noting that the bulk band gap in Bi$_{1-x}$Sb$_x$ is rather small ($\leq50$ meV) and its surface states is quite complex with multiple pieces of surface Fermi surfaces both near the $\bar{\Gamma}$ and the $\bar{M}$ points. Therefore, it is important to find a topological insulator consisting of a single surface state for the purposes of both studying their physical properties in fundamental physics and utilizing them in devices. This motivated a search for topological insulators with a larger band gap and simpler surface spectrum. A second generation of 3D topological insulator materials, especially Bi$_2$Se$_3$, offers the potential for topologically protected behavior in ordinary crystals at room temperature and zero magnetic field. Starting in 2008, work by the Princeton group used spin-ARPES and first-principles calculations to study the surface band structure of Bi$_2$Se$_3$ and observe the characteristic signature of a topological insulator in the form of a single Dirac cone that is spin-polarized (Fig.\ref{BiTe_spin}) such that it also carries a non-trivial Berry's phase \cite{Xia,Nature_2009}. Concurrent theoretical work by \cite{Zhang_nphys} used electronic structure methods to show that Bi$_2$Se$_3$ is just one of several new large band-gap topological insulators. These other materials were soon after also identified using this ARPES technique we describe \cite{YLChen,Hsieh_PRL}. The Bi$_2$Se$_3$ surface state is found from spin-ARPES and theory to be a nearly idealized single Dirac cone as seen from the experimental data in Fig. \ref{RT}. An added advantage is that Bi$_2$Se$_3$ is stoichiometric (i.e., a pure compound rather than an alloy such as Bi$_{1-x}$Sb$_x$) and hence can be prepared, in principle, at higher purity. While the topological insulator phase is predicted to be quite robust to disorder, many experimental probes of the phase, including ARPES of the surface band structure, are clearer in high-purity samples. Finally and perhaps most important for applications, Bi$_2$Se$_3$ has a large band gap of around 0.3 eV (3600 K). This indicates that in its high-purity form Bi$_2$Se$_3$ can exhibit topological insulator behavior at room temperature and greatly increases the potential for applications. Now, Bi$_2$Se$_3$ has become the prototype TI that features a single-Dirac-cone topological surface state, which is widely used for many transport, tunneling, optical, nanostructured studies.

\begin{figure*}
\includegraphics[width=14cm]{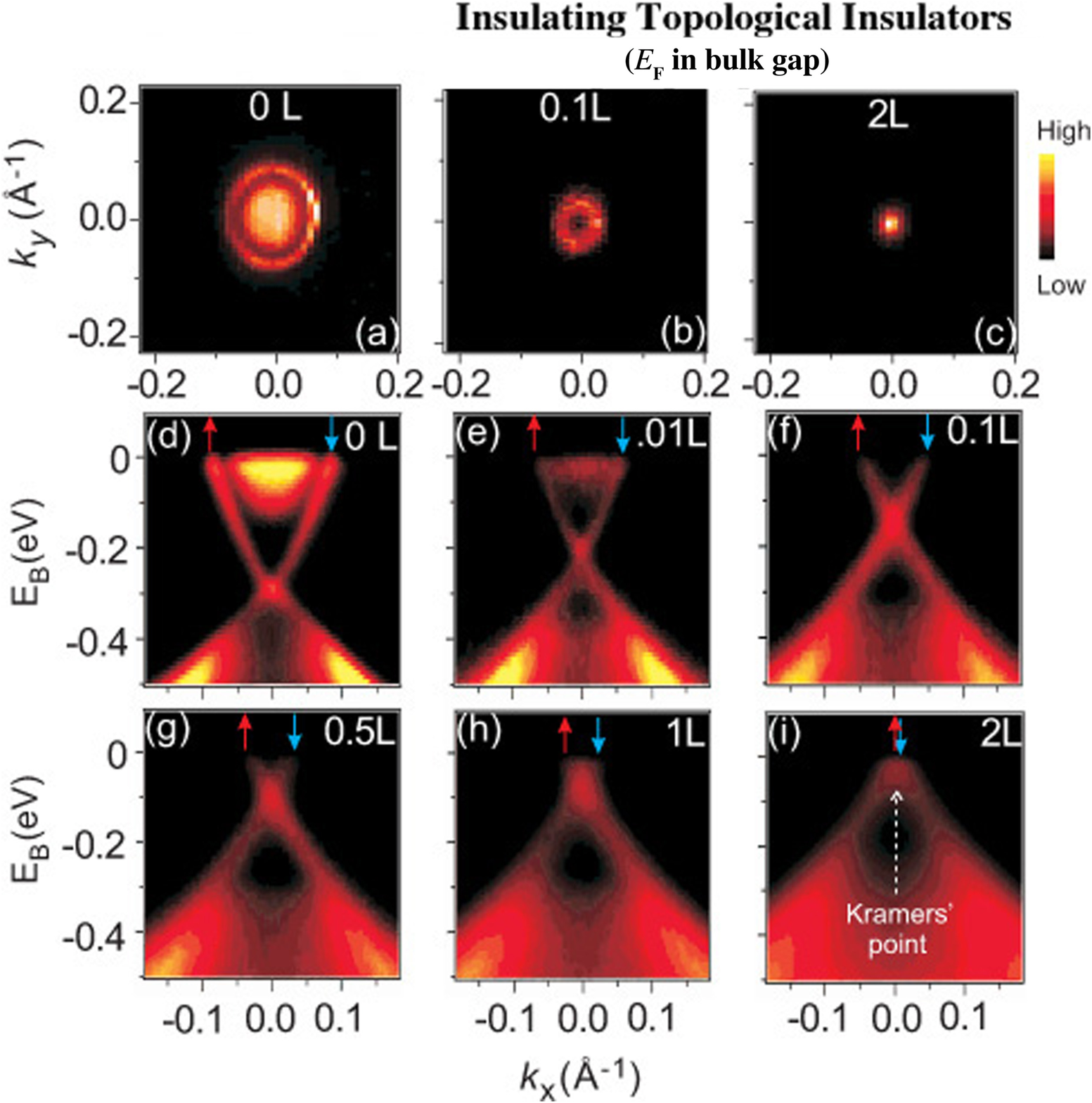}
\caption{\label{fig:NO2} \textbf{Surface gating : Tuning the density of helical Dirac electrons to the spin-degenerate Kramers point and topological transport regime.} (a) A high resolution ARPES mapping of the surface Fermi surface (FS) near $\bar{\Gamma}$ of Bi$_{2-\delta}$Ca$_{\delta}$Se$_3$ (111). The diffuse intensity within the ring originates from the bulk-surface resonance state \cite{15}. (b) The FS after 0.1 Langmuir (L) of NO$_2$ is dosed, showing that the resonance state is removed. (c) The FS after a 2 L dosage, which achieves the Dirac charge neutrality point. (d) High resolution ARPES surface band dispersions through after an NO$_2$ dosage of 0 L, (e) 0.01 L, (f) 0.1 L, (g) 0.5 L, (h) 1 L and (i) 2 L. The arrows denote the spin polarization of the bands. We note that due to an increasing level of surface disorder with NO$_2$ adsorption, the measured spectra become progressively more diffuse and the total photoemission intensity from the buried Bi$_{2-\delta}$Ca$_{\delta}$Se$_3$ surface is gradually reduced. [Adapted from D. Hsieh $et$ $al.$, \textit{Nature} \textbf{460}, 1101 (2009). \cite{Nature_2009}].}
\end{figure*}

Besides the Z$_2$ topological variant $\nu_0$, there is another topological number that can be uniquely determined by our spin-resolved ARPES measurements, the topological mirror Chern number $n_M$. For example, we now determine the value of $n_M$ of antimony surface states from our data. As shown in figure \ref{BiSb_spin}A, the vertical plane along the $\bar{\Gamma}-\bar{M}$ direction (yellow plane in figure \ref{BiSb_spin}A) is a mirror plane for the bulk BZ of antimony. Therefore, the electronic states within this mirror plane are the eigenstates of the Mirror operator, which defines a topological number that is the topological mirror Chern number $n_M$. The absolute value of the mirror Chern number $|n_M|$ is determined by the number of surface states moving to $+k$ (or $-k$) along the $\bar{\Gamma}-\bar{M}$ direction (the $\bar{\Gamma}-\bar{M}$ is the projection of the mirror plane onto the (111) surface). From figure \ref{BiSb_spin}, it is seen that a single (one) surface band, which switches partners at $\bar{M}$, connects the bulk valence and conduction bands, so $|n_M|$ = 1 . The sign of $n_M$ is related to the direction of the spin polarization $\langle \vec{P} \rangle$ of this band \cite{20}, which is constrained by mirror symmetry to point along $\pm\hat{y}$. Since the central electron-like FS enclosing $\bar{\Gamma}$ intersects six mirror invariant points, the sign of $n_M$ distinguishes two distinct types of handedness for this spin polarized FS. Figure \ref{BiSb_spin}(F) shows that for both Bi$_{1-x}$Sb$_x$ and Sb, the surface band that forms this electron pocket has $\langle \vec{P} \rangle \propto -\hat{y}$ along the $k_x$ direction, suggesting a left-handed rotation sense for the spins around this central FS thus $n_M$ = $-1$. We note that similar analysis regarding the mirror symmetry and mirror eigenvalues $n_M=-1$ can be applied to the single Dirac cone surface states in the Bi$_2$Se$_3$ material class. In fact, a nonzero (nontrivial) topological mirror Chern number does not require a nonzero Z$_2$ topological number. Or in other word, there is no necessary correlation between a mirror symmetry protected topological order and a time-reversal symmetry protected topological order. However, since most of the Z$_2$ topological insulators (Bi$_{1-x}$Sb$_x$, Bi$_2$Se$_3$, Bi$_2$Te$_3$ and etc.) also possess mirror symmetries in their crystalline form, thus topological mirror order $n_M$ is typically ``masked'' by the strong Z$_2$ topological order. One possible way to isolate the mirror topological order from the Z$_2$ order is to work with systems that feature an even number of bulk band inversions. This approach naturally exclude a nontrivial Z$_2$ order which strictly requires an odd number of band inversions.  More importantly, if the locations of the band inversions coincide with the mirror planes in momentum space, it will lead to a topologically nontrivial phase protected by the mirror symmetries of the crystalline system that is irrelevant to the time-reversal symmetry protection and the Z$_2$ (Kane-Mele) topological order. Such exotic new phase of topological order, noted as topological crystalline insulator \cite{Liang PRL TCI} protected by space group mirror symmetries, has very recently been theoretically predicted and experimentally identified in the Pb$_{1-x}$Sn$_x$Te(Se) alloy systems \cite{Liang NC SnTe, TCI Hasan, TCI Story, TCI Ando}. An anomalous $n_M=-2$ topological mirror number in Pb$_{1-x}$Sn$_x$Te, distinct from the $n_M=-1$ case observed in the Z$_2$ topological insulators, has also been experimentally determined using spin-resolved ARPES measurements as shown in Ref. \cite{TCI Hasan}. Moreover, the mirror symmetry can be generalized to other space group symmetries, leading to a large number of distinct topological crystalline insulators awaited to be discovered, some of which are predicted to exhibit nontrivial crystalline order even without spin-orbit coupling as well as topological crystalline surface states in non-Dirac (e.g. quadratic) fermion forms \cite{Liang PRL TCI}.

It can be seen from our ARPES data that as grown Bi$_2$Se$_3$ is in a doped semiconductor, where the chemical potential cuts into the bulk conduction band. The observed $n$-type behavior is believed to be caused by Se vacancies. However, many of the interesting theoretical proposals that utilize topological insulator surfaces require the chemical potential to lie at or near the surface Dirac point. This is similar to the case in graphene, where the chemistry of carbon atoms naturally locates the Fermi level at the Dirac point. This makes its density of carriers highly tunable by an applied electrical field and enables applications of graphene to both basic science and microelectronics. We demonstrated \cite{Nature_2009} that appropriate chemical modifications both in the bulk and on the surface of, which does not change its topologican nature, can achieve the condition with the chemical potential at the surface Dirac point, however, the Fermi energy of both the bulk and the surface can be controlled. This was achieved by doping bulk with a small concentration of Ca, which compensates the Se vacancies, to place the Fermi level within the bulk band gap. The surface was the hole doped by exposing the surface to NO$_2$ gas to place the Fermi level at the Dirac point, and has been shown to be effective even at room temperature (Fig.\ref{RT}). These results collectively show how ARPES can be used to study the topological protection and tunability properties of the 2D surface of a 3D topological insulator.

In summary, in this section, we have reviewed the experimental discovery of the 3D Z$_2$ topological insulator phase in Bi$_{1-x}$Sb$_x$, Bi$_2$Se$_3$ and other related compounds. Utilizing spin-resolved angle-resolved photoemission spectroscopy (spin-ARPES), we experimentally probed the nontrivial topological order ($\nu_0=1$) in 3D Z$_2$ TIs by measuring the surface state Fermi surface topology (number of surface state Fermi pockets enclosing the Kramers' points) and the spin-polarization texture that determines the $\pi$ quantum Berry's phase. Furthermore, we demonstrated that the topological surface states are realized and stable even at room temperatures, their chemical potential can be tuned and engineered to achieve the charge neutrality (Dirac) point, and furthermore their unique spin-momentum locking property leads to the prohibition of backscatting. These important experimental observations via spin-ARPES, not only demonstrate the 3D Z$_2$ topological insulator phase and its topological order ($\nu_0=1$), but also provide a powerful and unique methodology that is now used to discover and study new topological order in three-dimensions, such as the topological Kondo insulator, the topological crystalline insulator, the topological Dirac semimetal phases that we will discuss in the following sections.

\section{Topological Kondo Insulator Candidates}

Materials with strong electron correlations often exhibit exotic ground states such as the heavy fermion behavior, Mott or Kondo insulation and unconventional superconductivity. Kondo insulators are mostly realized in the rare-earth based compounds featuring \textit{f}-electron degrees of freedom, which behave like a correlated metal at high temperatures whereas a bulk bandgap opens at low temperatures through the hybridization \cite{Fisk_CMP, Coleman, Riseborough} of nearly localized-flat \textit{f} bands with the \textit{d}-derived dispersive conduction band. With the advent of topological insulators \cite{RMP, Hsieh1, FuKM, Xia} the compound SmB$_6$, often categorized as a heavy-fermion semiconductor \cite{Fisk_CMP, Coleman, Riseborough}, attracted much attention due to the proposal that it may possibly host a topological Kondo phase (TKI) at low temperatures where transport is anomalous \cite{Dzero,Takimoto, Dai_SmB6}. The anomalous residual conductivity is believed to be associated with electronic states that lie within the Kondo gap \cite{Kimura, Menth, Allen, Cooley, Nanba, Nyhus, Alekseev, point_cont, Miyazaki, Denlinger}.

Following the prediction of a TKI phase, there have been several surface-sensitive transport measurements, which include observation of a three-dimensional (3D) to two-dimensional (2D) crossover of the transport carriers below $T\sim 7 $K \cite{Fisk, Hall, Fisk2}. However, due to the lack of the critical momentum resolution for the transport probes, neither the existence of in-gap surface states nor their Fermi surface topology (number of surface Fermi surfaces and enclosing or not enclosing the Kramer's points) have been experimentally studied. By combining high-resolution laser- and synchrotron-based angle-resolved photoemission techniques in Ref. \cite{MN}, we present the surface electronic structure identifying the in-gap states that are strongly temperature dependent and disappear before approaching the coherent Kondo hybridization scale. Remarkably, the observed Fermi surface for the low-energy part of the in-gap states keeping the sample within the transport anomaly regime ($T\sim6$ K) reveals an odd number of pockets that enclose three out of the four Kramers' points of the surface Brillouin zone, consistent with the theoretically calculated Fermi surface topology of the topological surface states. Concurrent ARPES studies on SmB$_6$ are also reported in Refs. \cite{Feng, Shi}.

\begin{figure*}
\centering
\includegraphics[width=16.5cm]{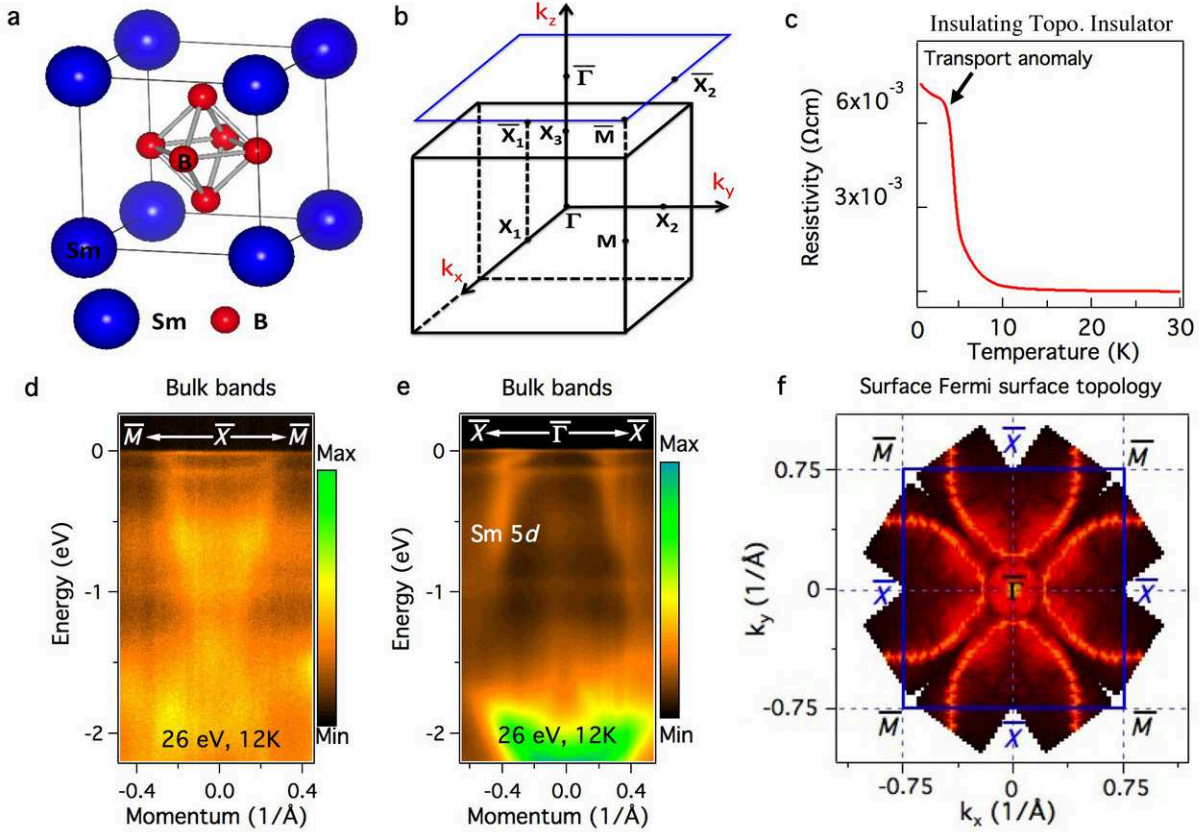}
\caption{\label{SmB6_1}\textbf{Brillouin zone symmetry, and band structure of SmB$_6$}. \textbf{a,} Crystal structure of SmB$_6$. Sm ions and B$_6$ octahedron are located at the corners and the center of the cubic lattice structure. \textbf{b,} The bulk and surface Brillouin zones of SmB$_6$. High-symmetry points are marked. \textbf{c,} Resistivity-temperature profile for samples used in ARPES measurements. \textbf{d, e,} Synchrotron-based ARPES dispersion maps along the ${\bar{M}}-{\bar{X}}-{\bar{M}}$ and the ${\bar{X}}-{\bar\Gamma}-{\bar{X}}$ momentum-space cut-directions. Dispersive Sm 5$d$ band and non-dispersive flat Sm 4$f$ bands are observed, confirming the key ingredient for a heavy fermion Kondo system. \textbf{f,} A Fermi surface map of bulk insulating SmB$_6$ using a 7 eV laser source at a sample temperature of $\simeq6$ K (Resistivity$=5$ m$\Omega$cm), obtained within the $E_{\textrm{F}}\pm4$ meV window, which captured all the low energy states between 0 to 4 meV binding energies, where in-gap surface state's spectral weight contribute most significantly within the insulating Kondo gap . Intensity contours around $\bar{\Gamma}$ and ${\bar{X}}$ reflect low-lying metallic states near the Fermi level, which is consistent with the theoretically predicted Fermi surface \textit{topology} of the topological surface states. [Adapted form M. Neupane \textit{et al.,} \textit{Nature Commun.} \textbf{4}, 2991 (2013)].}
\end{figure*}

SmB$_6$ crystallizes in the CsCl-type structure with the Sm ions and the B$_6$ octahedra being located at the corner and at the body center of the cubic lattice, respectively (Fig.~\ref{SmB6_1}\textbf{a}). The bulk Brillouin zone (BZ) is a cube made up of six square faces. The center of the cube is the $\Gamma$ point, whereas the centers of the square faces are the $X$ points. Due to the inversion symmetry of the crystal, each $X$ point and its diametrically opposite partner are completely equivalent. Therefore, there exist three distinct $X$ points in the BZ, labeled as $X_1$, $X_2$ and $X_3$. It is well-established that the low energy physics in SmB$_6$ is constituted of the non-dispersive Sm $4f$ band and the dispersive Sm $5d$ band located near the $X$ points \cite{Fisk, Hall, tunelling, Miyazaki, Denlinger}. Figs.~\ref{SmB6_1}\textbf{d} and \textbf{e} show ARPES intensity profiles over a wide binding energy scale measured with a synchrotron-based ARPES system using a photon energy of 26 eV. The dispersive features originate from the Sm $5d$ derived bands and a hybridization between the Sm $5d$ band and Sm 4$f$ flat band is visible especially around 150 meV binding energies confirming the Kondo features of the electronic system in our study (Figs.~\ref{SmB6_1}\textbf{d} and \textbf{e}).

\begin{figure*}
\centering
\includegraphics[width=14cm]{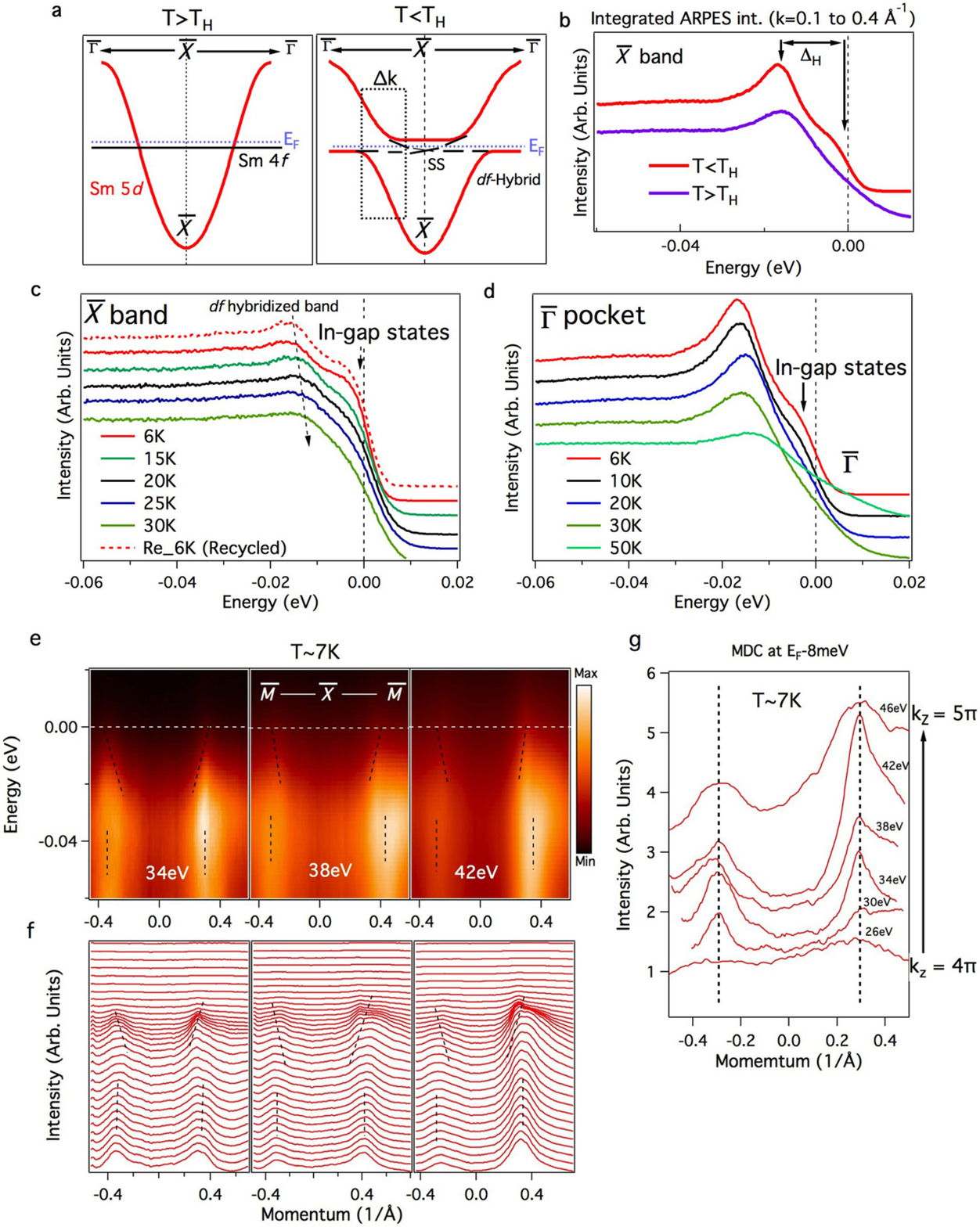}
\caption{\label{SmB6_2}\textbf{Temperature dependent in-gap states and its two-dimensional nature.} \textbf{a,} Cartoon sketch depicting the basics of Kondo lattice hybridization at temperatures above and below the hybridization gap opening. The blue dashed line represents the Fermi level in bulk insulating samples such as SmB$_6$ (since the bulk of the SmB$_6$ is insulating, the Fermi level must lie within the Kondo gap). The theoretically predicted topological surface states within the Kondo gap are also shown in this cartoon view (black dash lines) based on Refs \cite{Takimoto, Dai_SmB6}. The black dash rectangle shows the approximate momentum window of our laser-ARPES measurements between $k_1=0.1$ $\textrm{\AA}^{-1}$ to $k_2=0.4$ $\textrm{\AA}^{-1}$. \textbf{b,} Partially momentum-integrated ARPES spectral intensity in a $\pm0.15$ $\textrm{\AA}^{-1}$ window ($\Delta{k}$ defined in panel \textbf{a}) above and below the Kondo lattice hybridization temperature ($T_{\textrm{H}}$). \textbf{c,} Momentum-integrated ARPES spectral intensity centered at the $\bar{X}$ point at various temperatures. \textbf{d,} Analogous measurements as in Panel \textbf{c} but centered at the $\bar{\Gamma}$ pocket ($\Delta{k}=0.3$ $\textrm{\AA}^{-1}$). ARPES data taken on the sample after thermally recycling (6 K up to 50 K then back to 6 K) is shown by Re$\_$6K, which demonstrates that the in-gap states are robust against thermal recycling. \textbf{e,} Synchrotron based ARPES energy momentum dispersion maps measured using different photon energies along the ${\bar{M}}-{\bar{X}}-{\bar{M}}$ momentum space cut-direction. Incident photon energies used are noted on the plot. \textbf{f,} Momentum distribution curves (MDCs) of data shown in \textbf{a}. The peaks of the momentum distribution curves are marked by dashed lines near the Fermi level, which track the dispersion of the low-energy states. \textbf{g,} Momentum distribution curves in the close vicinity of the Fermi level (covering the in-gap states near the gap edge) integrated within the energy window of [$E_{\textrm{F}}$ - 8 meV, $E_{\textrm{F}}$] are shown as a function of photon energy which covers the $k_z$ range of $4\pi$ to $5\pi$ at 7 K. [Adapted form M. Neupane \textit{et al.,} \textit{Nature Commun.} \textbf{4}, 2991 (2013)].}
\end{figure*}

In order to search for the predicted in-gap states within 5 meV of the Fermi level, a laser-based ARPES system providing $\Delta{E}\sim4$ meV coupled with a low temperature ($T\simeq5$ K) capability is employed in Ref. \cite{MN}. Since the low-energy physics including the Kondo hybridization process occurs near the three $X$ points (Fig.~\ref{SmB6_1}\textbf{f}) in the bulk BZ and the $X$ points project onto the $\bar{X}_1$, $\bar{X}_2$, and the $\bar{\Gamma}$ points at (001) surface (Fig.~\ref{SmB6_1}\textbf{b}), the Kramers' points of this lattice are $\bar{X}_1$, $\bar{X}_2$, $\bar{\Gamma}$ and $\bar{M}$ and one needs to systematically study the connectivity (winding) of the in-gap states around these points. Fig.~\ref{SmB6_2}\textbf{c} shows experimentally measured ARPES spectral intensity integrated in a narrow ($\pm0.15$ $\textrm{\AA}^{-1}$) momentum window and their temperature evolution around the $\bar{X}$ point. At temperatures above the hybridization scale, only one spectral intensity feature is observed around $E_{\textrm{B}}\sim12$ meV in the ARPES EDC profile. As temperature decreases below 30 K, this feature is found to move to deeper binding energies away from the chemical potential, consistent with the opening of the Kondo hybridization gap while Fermi level is in the insulating gap (bulk is insulating, according to transport, so Fermi level must lie in-gap at 6 K). At lower temperatures, the gap value of hybridized states at this momentum space regime is estimated to be about 16 meV. More importantly, at a low temperature $T\simeq6$ K corresponding to the 2D transport regime, a second spectral intensity feature is observed at the binding energy of $E_{\textrm{B}}\sim4$ meV, which lies inside the insulating gap. Our data thus experimentally shows the existence of in-gap states. Remarkably, the in-gap state feature is most pronounced at low temperature $T\simeq6$ K in the 2D transport regime, but becomes suppressed and eventually vanishes as temperature is raised before reaching the onset for the Kondo lattice hybridization at 30 K. The in-gap states are found to be robust against thermal cycling, since lowering the temperature back down to 6K results in the similar spectra with the re-appearance of the in-gap state features (Re\_6K in Fig.~\ref{SmB6_2}\textbf{c}). The observed robustness against thermal recyclings counts against the possibility of non-robust (trivial) or non-reproducible surface states. We further performed similar measurements of low-lying states focusing near the $\bar{\Gamma}$ point (projection of the $X_3$) as shown in Fig.~\ref{SmB6_2}\textbf{d}. Similar spectra reveal in-gap state features prominently around $E_{\textrm{B}}\sim3-4$ meV at $T\simeq6$ K which clearly lie within the Kondo gap and exhibit similar (coupled) temperature evolution as seen in the spectra obtained near the $\bar{X}$ point.


We further study their momentum-resolved structure or the k-space map for investigations regarding their topology: 1) The number of surface state pockets that lie within the Kondo gap; 2) The momentum space locations of the pockets (whether enclosing or winding the Kramers' points or not). Fig.~\ref{SmB6_1}\textbf{f} shows a Fermi surface map measured by setting the energy window to cover $E_{\textrm{F}}\pm4$ meV, which ensures the inclusion of the in-gap states (that show temperature dependence consistent with coupling to the Kondo hybridization) within the Fermi surface mapping data as identified in Fig.~\ref{SmB6_2}\textbf{d,e} at a temperature of 6 K inside the 2D transport anomaly regime under the ``better than 5 meV and 7 K combined resolution condition''. Our Fermi surface mapping reveals multiple pockets which consist of an oval-shaped as well as nearly circular-shaped pockets around the $\bar{X}$ and $\bar{\Gamma}$ points, respectively. No pocket was seen around the $\bar{M}$-point which was measured in a synchrotron ARPES setting. Therefore the laser ARPES data captures all the pockets that exist while the bulk is insulating. This result is striking by itself from the point of view that while we know from transport that the bulk is insulating, ARPES shows large Fermi surface pockets (metallicity of the surface) at this temperature. Another unusual aspect is that not all Kramers' points are enclosed by the in-gap states. Our observed Fermi surface thus consists of 3 (or odd number Mod 2 around each Kramers' point) pockets per Brilluoin zone and each of them wind around a Kramers' point only and this number is odd (at least 3). Therefore, our measured in-(Kondo) gap states lead to a very specific form of the Fermi surface topology (Fig.~\ref{SmB6_1}\textbf{f}) that is remarkably consistent with the theoretically predicted topological surface state Fermi surface expected in the TKI groundstate phase despite the broad nature of the contours.

Since for the laser-ARPES, the photon energy is fixed (7 eV) and the momentum window is rather limited (the momentum range is proportional to $\sqrt{h\nu-W}$, where $h\nu$ is the photon energy and $W\simeq4.5$ eV is the work function), we utilize synchrotron based ARPES measurements to study the low-lying state as a function of photon energy as demonstrated in Bi-based topological insulators \cite{RMP}. Fig.~\ref{SmB6_2}\textbf{e,f} show the energy-momentum cuts measured with varying photon energies. Clear $E-k$ dispersions are observed within a narrow energy window near the Fermi level. The dispersion is found to be unchanged upon varying photon energy, supporting their quasi-two-dimensional nature (see, Fig.~\ref{SmB6_2}\textbf{g}). The observed quasi-two-dimensional character of the signal within 10 meV of the gap where surface states reside does suggest consistency with the surface nature of the in-gap states. Due to the combined effects of energy resolution ($\Delta{E}\geq10$ meV, even though the sample temperature, 7 K, is near the anomalous transport regime) and the intrinsic self-energy broadening coupled with the higher weight of the \textit{f}-part of the cross-section and the strong band tails, the in-gap states are intermixed with the higher energy bulk bands' tails.
In order to isolate the in-gap states from the bulk band tails that have higher cross-section at synchrotron photon energies, it is necessary to have energy resolution (not just the low working temperature) better than half the Kondo gap scale which is about 7 meV or smaller in SmB$_6$.
Our experiment reports of Fermi surface mapping covering the low-energy part of the in-gap states keeping the sample within the transport anomaly regime reveals an odd number of pockets that enclose three out of the four Kramers' points of the surface Brillouin zone strongly suggesting the the topological origin of the in-gap state.

\section{Topological Quantum Phase Transitions}

A three-dimensional topological insulator is a new phase of matter distinct from a conventional band insulator (semiconductor) in that a TI features a nontrivial topological invariant in its bulk electronic wavefunction space \cite{RMP, Moore1, HasanMoore, FuKM, Roy, 15, Hsieh1, 10, Science, Xia, Murakami, Pallab, Gil, Qi,Chiral, Galitski, SUSY}. The nonzero topological invariant in a TI leads to the existence of spin-momentum locked gapless Dirac electrons on its surfaces \cite{Hsieh1, 10, Science, Xia}. It has been theoretically known that a TI can be tuned from a conventional insulator by going through an adiabatic band inversion process in the bulk  \cite{RMP, FuKM, Murakami}. Such a quantum phase transition from a conventional band insulator to a TI that involves a change of the bulk topological invariant is defined as a topological phase transition. The topological phase transition is of great interest because its critical point (the topological-critical-point) is expected to not only realize new groundstates such as higher dimensional Dirac/Weyl fermions \cite{Murakami, Pallab, Gil, Qi} and supersymmetry state \cite{SUSY}, but also show exotic transport and optical responses such as chiral anomaly in magnetoresistence \cite{Chiral} or the light-induced Floquet topological insulator state \cite{Galitski}. To achieve these novel phenomena, it is of importance to study the electronic and spin groundstate across a topological phase transition. Studying the electronic and spin groundstate across a topological phase transition also serves as the key to understanding the formation of the topological surface states across the topological phase transition. It is well established that the topological Dirac surface states and their spin-momentum locking are the signature that distinguishes a topological insulator from a conventional insulator. However, an interesting and vital question that remains unanswered is how topological surface states emerge as a non-topological system approaches and crosses the topological criticality. Therefore, in order to realize these proposed new topological phenomena and also to understand the fate of the topological surface states across the topological-critical-point, it is critically important to realize a fully tunable spin-orbit real material system, where such topological phase transition can be systematically realized, observed, and further engineered.

Such a fully tunable topological phase transition system is first realized by our ARPES and spin-resolved ARPES studies on the BiTl(S$_{1-\delta}$Se$_\delta$)$_2$ \cite{Xu}. In Ref. \cite{Xu}, by studying the electronic and spin groundstate of the BiTl(S$_{1-\delta}$Se$_\delta$)$_2$ samples with various $\delta$ compositions, a bulk band inversion and a topological phase transition between a conventional band insulator and a topological insulator is, for the first time, systematically demonstrated and visualized. Such study \cite{Xu} serves as a corner stone for realizing new topological phenomena based on the topological phase transition as discussed above \cite{Murakami, Pallab, Gil, Qi,Chiral, Galitski, SUSY}. This work was therefore followed and expanded by many later works (e.g. \cite{Ando QPT, Oh, Preform, BiTeI}), which not only studied the BiTl(S$_{1-\delta}$Se$_\delta$)$_2$ system in greater details and depth \cite{Ando QPT, Preform} but also expanded the realization of the topological phase transition into other classes of topological materials \cite{Oh, BiTeI}.

\begin{figure*}
\includegraphics[width=17cm]{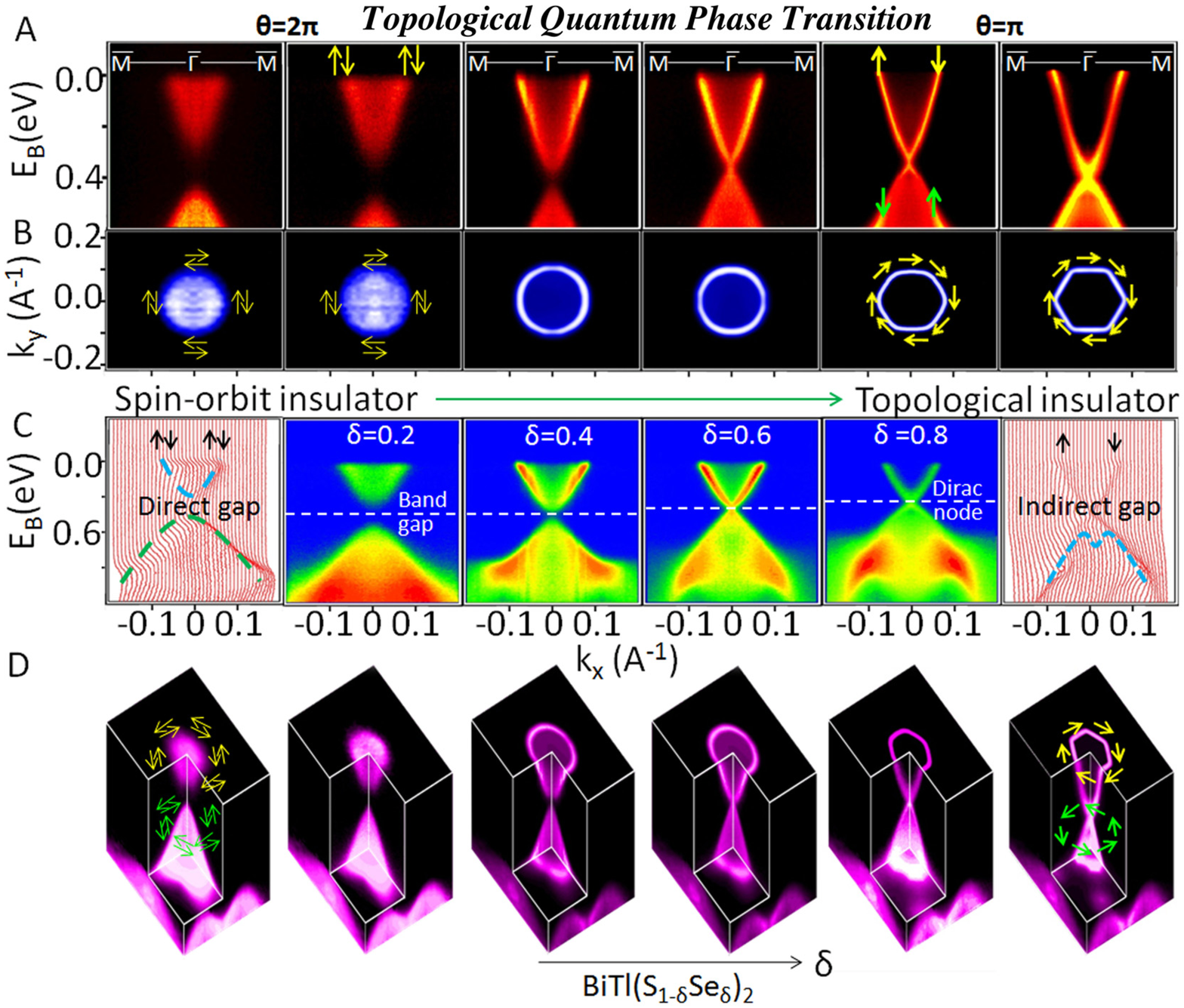}
\caption{\label{QPT1} \textbf{Topological phase transition through a 3D Dirac Semimetal (3D analog of graphene) phase in BiTl(S$_{1-{\delta}}$Se$_{\delta}$)$_2$.} \textbf{(A)} High resolution ARPES dispersion mappings along a pair of time-reversal invariant points or Kramers' points ($\bar{\Gamma}$ and $\bar{M}$). \textbf{(B)} ARPES mapped native Fermi surfaces for varying chemical compositions. \textbf{(C)} Left- and right-most: Energy-distribution curves for $\delta=0.0$ and 1.0. Middle: ARPES spectra with Fermi levels (white dotted lines) placed at the center of the band-gap or on the Dirac node, which highlights the evolution of the dispersion of valence band across the topological transition. In the middle 4 panels, the intensities above the white dotted lines are set to zero, as a guide to the eye. \textbf{(D)} Compositional evolution of band structure measured over a wide energy and momentum range. At the critical point a 3D Dirac Semimetal (3D analog of graphene) is realized [Adapted from S.-Y. Xu $et$ $al.$, \textit{Science} \textbf{332} 560 (2011). \cite{Xu}], Also see, M. Neupane \textit{et. al.,} \textit{Nature Commun.} (2014) for the critical point data on 3D Dirac Semimetal \cite{CdAs_Hasan}]}
\end{figure*}

\begin{figure*}
\includegraphics[width=17cm]{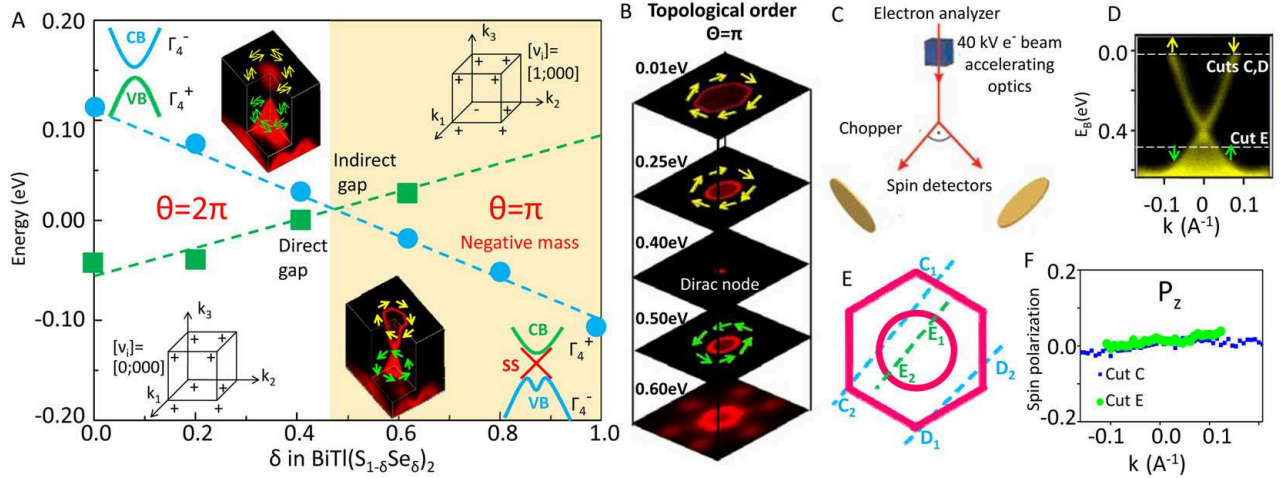}
\caption{\label{QPT2} \textbf{Bulk band inversion and spin texture inversion in BiTl(S$_{1-{\delta}}$Se$_{\delta}$)$_2$.} \textbf{(A)} High resolution ARPES measured Bi$_2$Te$_3$ Fermi surface. Direction and relative position of the spin-polarization measurement cuts A and B are indicated. \textbf{(B)} Out-of-plane spin-polarization profile of Bi$_2$Te$_3$ for cuts A and B. \textbf{(C)} Fitted values of direction of the 3D spin vectors obtained from the Bi$_2$Te$_3$ spin-resolved data. A 3D modulated spin texture is revealed from the data. \textbf{(D)} Experimental geometry employed to obtain the spin-polarization components. \textbf{(E)} ARPES measured BiTl(S$_{0}$Se$_{1}$)$_2$ dispersion along $\bar{\Gamma}-\bar{M}$ momentum space cut, indicating the energy positions of cuts C, D and E. The binding energies for the cuts are: $E_{\textrm{B}}$(Cuts C,D)=0.01eV, $E_{\textrm{B}}$(Cuts E)=0.50eV. \textbf{(F)} Measured out-of-plane spin-polarization profile of cuts C and E on BiTl(S$_{0}$Se$_{1}$)$_2$. \textbf{(G)} A map of the momentum space spin-resolved cuts C, D and E across the Fermi surfaces of BiTl(S$_{0}$Se$_{1}$)$_2$. The hexagonal Fermi surface is located 0.40eV above the Dirac node, whereas the circular Fermi surface is located 0.10eV below the Dirac node. \textbf{(H)} Surface Fermi surface topology evolution of BiTl(S$_{0}$Se$_{1}$)$_2$ across the Dirac node. The corresponding binding energies of constant energy contours are indicated. Observed spin textures are schematically drawn at various binding energies. [Adapted from S.-Y. Xu $et$ $al.$, \textit{Science} \textbf{332} 560 (2011). \cite{Xu}]}
\end{figure*}

Figure~\ref{QPT1}A presents systematic photoemission measurements of electronic states that lie between a pair of time-reversal invariant points or KramersÕ points ($\bar{\Gamma}$ and $\bar{M}$) obtained for a series of compositions of the spin-orbit material BiTl(S$_{1-\delta}$Se$_\delta$)$_2$. As the selenium concentration is increased, the low-lying bands separated by a gap of energy $0.15$ eV at $\delta=0.0$ are observed to approach each other and the gap decreases to less than $0.05$ eV at $\delta=0.4$. The absence of surface states (SSs) within the bulk gap suggests that the compound is topologically trivial for composition range of $\delta=0.0$ to $\delta=0.4$. Starting from $\delta=0.4$, a linearly dispersive band connecting the bulk conduction and valence bands emerges which threads across the bulk band gap. Moreover, the Dirac-like bands at $\delta=0.6$ and beyond are spin polarized (see Fig.~\ref{QPT2} ). The system enters a topologically non-trivial phase upon the occurrence of an electronic transition between $\delta=0.4$ and $\delta=0.6$. While the system approaches the transition from the conventional or no-surface-state side ($\delta=0.6$), both energy dispersion and FS mapping (Fig.~\ref{QPT1}A and B for $\delta=0.4$) show that the spectral weight at the outer boundary of the bulk conduction band continuum which corresponds to the loci where the Dirac SSs would eventually develop becomes much more intense; however, that the surface remains gapped at $\delta=0.4$ suggests that the material is still on the trivial side. A critical signature of a topological transition is that the material turns into an indirect bulk band gap material. As $\delta$ varies from 0.0 to 1.0 (Fig.~\ref{QPT1}C), the dispersion of the valence band evolves from a ``$\Lambda$''-shape to an ``$M$''-shape with a ``dip'' at the $\bar{\Gamma}$ point ($k=0$); the $\delta=0.0$ compound features a direct band gap in its bulk, whereas the $\delta=1.0$ indicates a slightly indirect gap. These systematic studies demonstrate the existence of the bulk band inversion and the topological phase transition between a conventional band insulator and a topological insulator in the BiTl(S$_{1-\delta}$Se$_\delta$)$_2$ system. The bulk band inversion process in the BiTl(S$_{1-\delta}$Se$_\delta$)$_2$ system is shown in Fig.~\ref{QPT2}.

\section{Topological Dirac Semimetals}

\begin{figure}
\centering
\includegraphics[width=9cm]{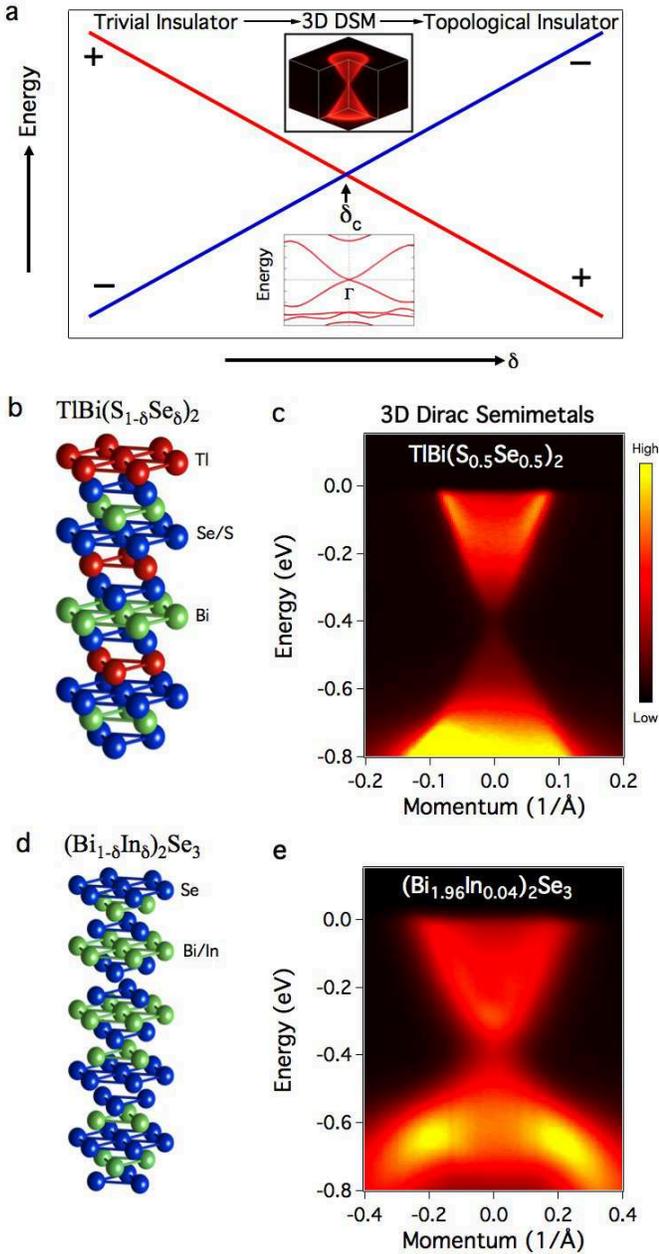}
\caption{\label{CdAs_0}\textbf{The 3D Dirac semimetal phase realized by fine-tuning to the critical point of a topological phase transition.} \textbf{a,} Schematic view of topological phase transition. The critical point ($\delta_c$) is marked by an arrow and a 3D Dirac cone is presented in the upper inset. The calculated electronic bulk bands structure [energy (eV) versus momentum (\AA$^{-1}$)]  at critical point is shown in the lower inset. \textbf{b,} Crystal structure of TlBi(S/Se)$_2$ with repeating
Tl-Se-Bi-Se layers. \textbf{c,} ARPES dispersion map of TlBi(S$_{1-\delta}$Se$_\delta$)$_2$ ($\delta=0.5$)
\textbf{d,}  Crystal structure of (Bi/In)$_2$Se$_3$ with repeating Bi/In-Se layers. \textbf{e,} ARPES dispersion map of (Bi$_{1-\delta}$In$_{\delta}$)$_2$Se$_3$ ($\delta=0.04$). Adapted from M. Neupane \textit{et. al.,} \textit{Nature Commun.} (2014) \cite{CdAs_Hasan}.}
\end{figure}

\begin{figure*}
\centering
\includegraphics[width=15cm]{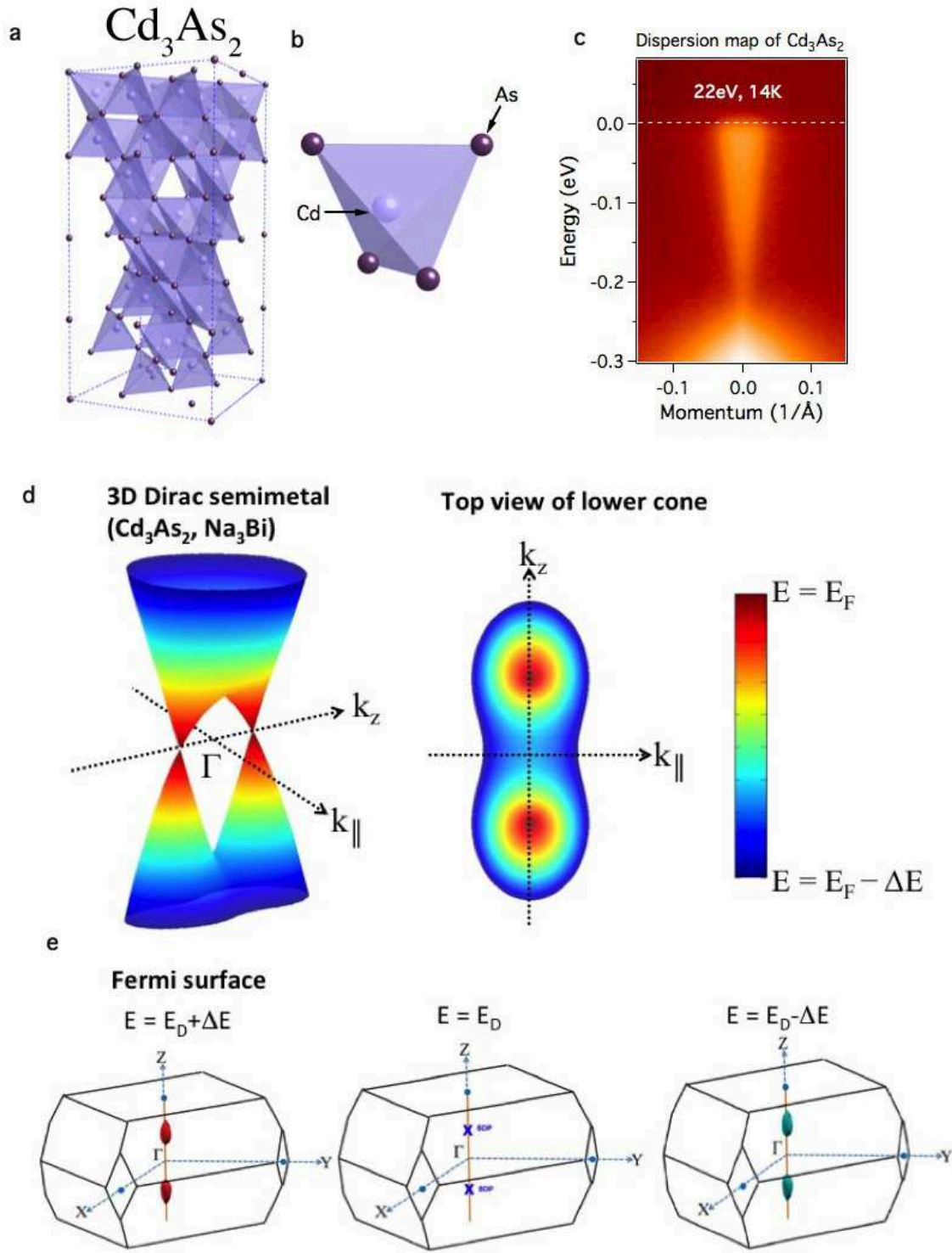}
\caption{\label{CdAs_1}\textbf{Topological Dirac semimetal phase in Cd$_3$As$_2$.} \textbf{a,} Cd$_3$As$_2$ crystalizes in a tetragonal body center structure with space group of $I4_1$cd, which has 32 number of formula units in the unit cell. The tetragonal structure has lattice constant of  $a= 12.670$ \AA, $b=12.670$ \AA, and $c=25.480$ \AA.}
\end{figure*}
\addtocounter{figure}{-1}
\begin{figure} [t!]
\caption{\textbf{b,} The basic structure unit is a 4 corner-sharing CdAs$_3$-trigonal pyramid. \textbf{c,} ARPES $E_{\textrm{B}}-k_x$ cut of Cd$_3$As$_2$ near the Fermi level at around surface BZ center $\bar{\Gamma}$ point. \textbf{d,} Cartoon view of dispersion of 3D Dirac semimetal. \textbf{e,} Schematic view of the Fermi surface above the Dirac point (left panel), at the Dirac point (middle panel) and below the Dirac point (right panel). [Figures are adapted from M. Neupane \textit{et. al.,} \textit{Nature Commun.} (2014) \cite{CdAs_Hasan}].}
\end{figure}

\begin{figure*}
\centering
\includegraphics[width=15cm]{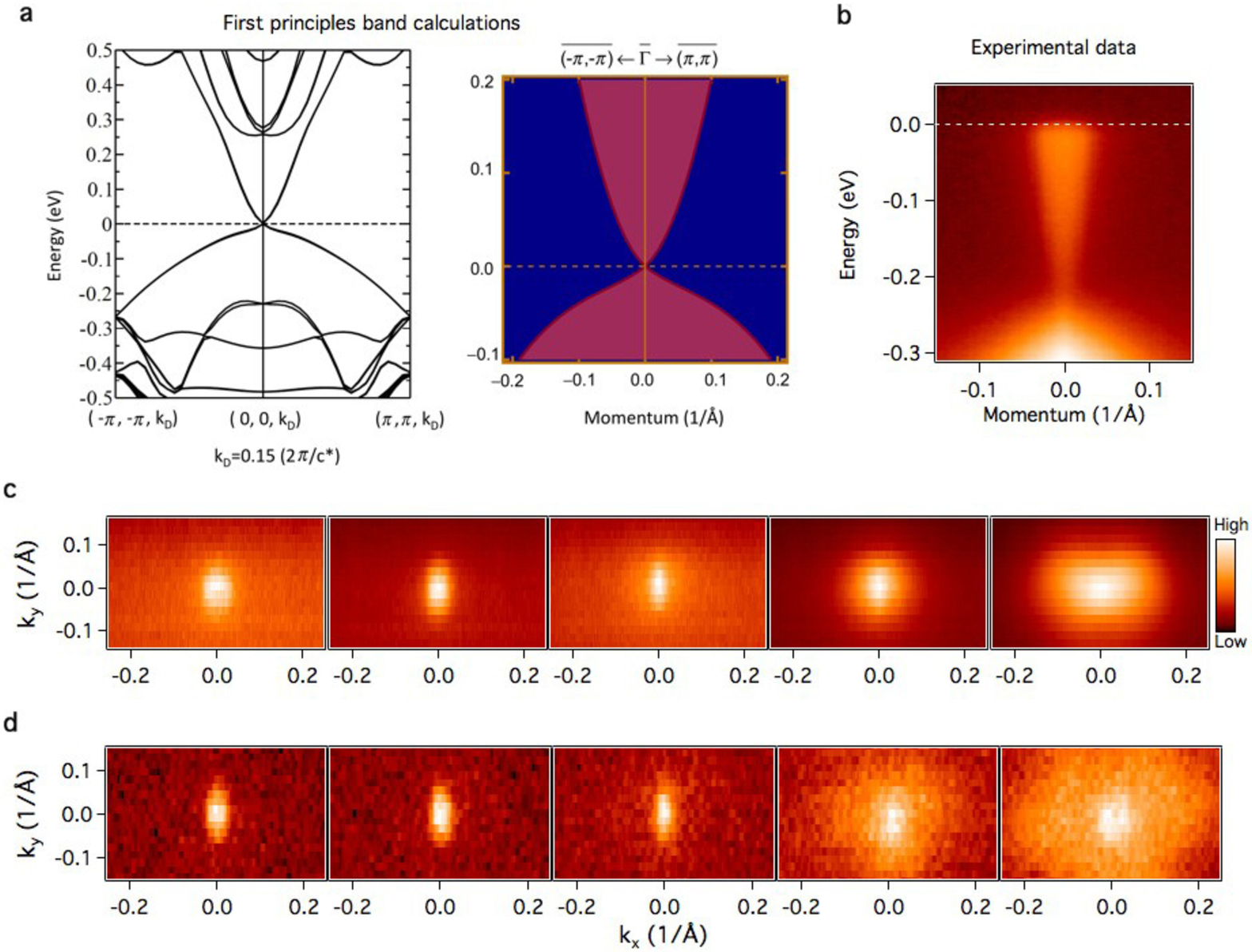}
\caption{\label{CdAs_2}\textbf{In-plane dispersion in Cd$_3$As$_2$.} \textbf{a,} Left: First principles calculation of the bulk electronic structure along the $(\pi,\pi,0.15\frac{2\pi}{c*})-(0,0,0.15\frac{2\pi}{c*})$ direction ($c*=c/a$). Right: Projected bulk band structure on to the (001) surface, where the shaded area shows the projection of the bulk bands. \textbf{b,} ARPES measured dispersion map of Cd$_3$As$_2$, measured with photon energy of 22 eV and temperature of 15 K along the $(-\pi,-\pi)-(0,0)-(\pi,\pi)$ momentum space cut direction. \textbf{c,} ARPES constant energy contour maps using photon energy of 22 eV on Cd$_3$As$_2$. \textbf{d,}  ARPES constant energy contour maps using photon energy of 102 eV on Cd$_3$As$_2$. [This figure is adapted from M. Neupane \textit{et al.,} \textit{Nature Commun.} (2014) \cite{CdAs_Hasan}]}
\end{figure*}

\begin{figure*}
\centering
\includegraphics[width=15cm]{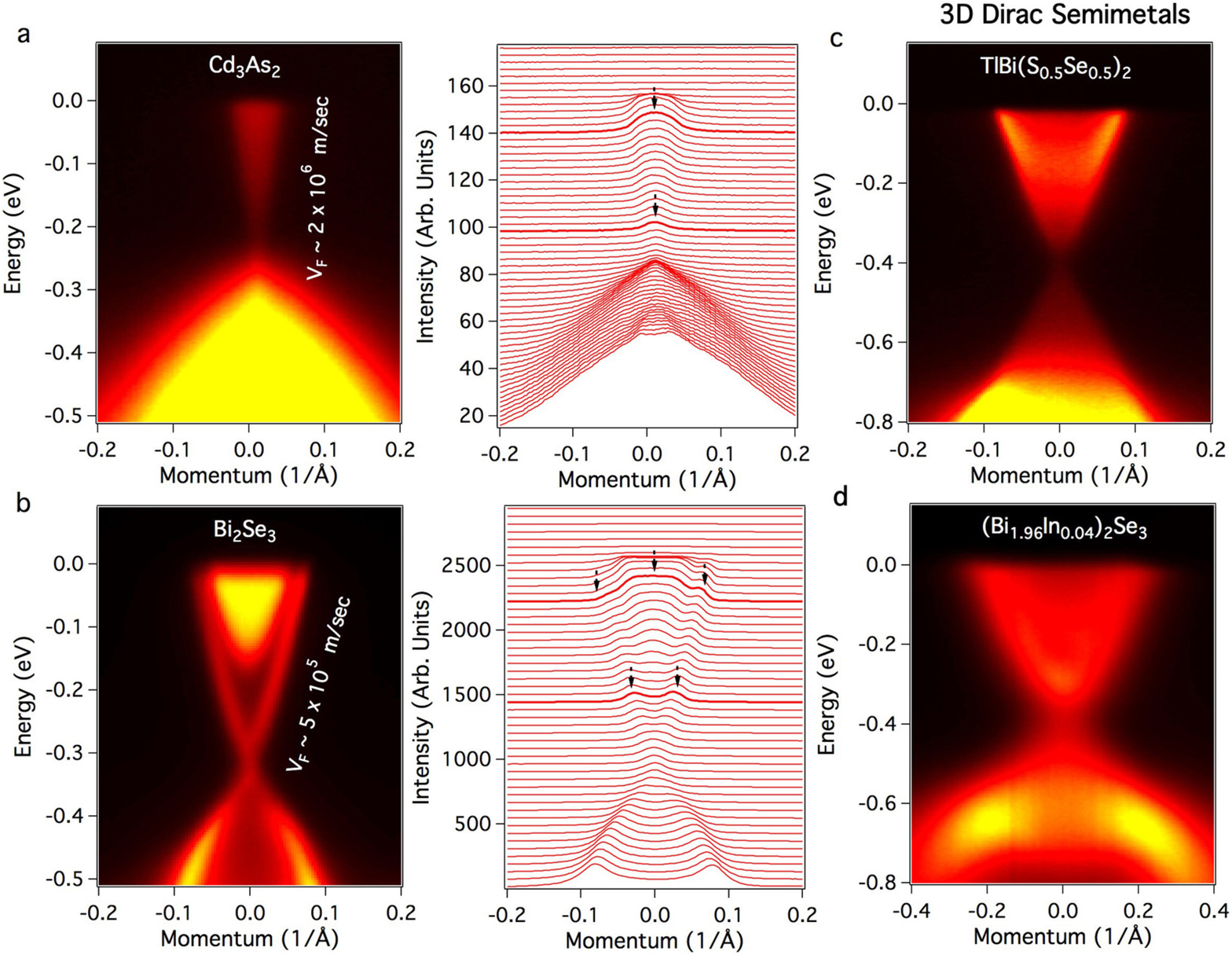}
\caption{\label{CdAs_3}\textbf{Surface electronic structure of 2D and 3D Dirac fermions.} \textbf{a,} ARPES measured surface electronic structure dispersion map of Cd$_3$As$_2$ and its corresponding momentum distribution curves (MDCs). \textbf{b,} ARPES measured surface dispersion map of the prototype TI Bi$_2$Se$_3$ and its corresponding momentum distribution curves. Both spectra are measured with photon energy of 22 eV and at a sample temperature of 15 K. The black arrows show the ARPES intensity peaks in the MDC plots. \textbf{c and d} ARPES spectra of two Bi-based 3D Dirac semimetals, which are realized by fine tuning the chemical composition to the critical point of a topological phase transition between a normal insulator and a TI: \textbf{c,} TlBi(S$_{1-\delta}$Se$_\delta$)$_2$ ($\delta=0.5$) (Xu $et$ $al.$ \cite{Xu}), and (Bi$_{1-\delta}$In$_{\delta}$)$_2$Se$_3$ ($\delta=0.04$) (Brahlek $et$ $al.$ \cite{Oh}) \textbf{d,}. Spectrum in panel \textbf{c} is measured with photon energy of 16 eV and spectrum in panel \textbf{d} is measured with photon energy of 41 eV. For the 2D topological surface Dirac cone in Bi$_2$Se$_3$, a distinct in-plane ($E_{\textrm{B}}-k_x$) dispersion is observed in ARPES, whereas for the 3D bulk Dirac cones in Cd$_3$As$_2$, TlBi(S$_{0.5}$Se$_{0.5}$)$_2$, and (Bi$_{0.96}$In$_{0.04}$)$_{2}$Se$_3$, a Dirac-cone-like intensity continuum is also observed. [This figure is adapted from M. Neupane \textit{et al.,} \textit{Nature Commun.} (2014) \cite{CdAs_Hasan}]}
\end{figure*}

The relativistic (Dirac) fermions of solid-state band structure has been known since 1947 \cite{Wallace} where graphene is considered. Graphene is a two dimensional semimetal and its electrons are effectively relativistic with velocity $(1/300)c$, where $c$ is the velocity of light in vacuum. The Dirac nature of the band structure is protected by symmetries of the graphene lattice. Recent realizations of two-dimensional massless Dirac electrons in  graphene and surfaces of the 3D topological insulator have generated enormous interest in condensed matter physics \cite{Graphene, Kim, Weyl, RMP, Hsieh1, Xia, Dirac_3D, Bismuth, Dirac_semi, Murakami, Dai, Volovik, Fang, Wan, Balent}. Many interesting phenomena such as exotic integer quantum Hall effect \cite{Kim} has been observed in graphene. It is known that a minimum model for a 2D Dirac electronic system is $H = v_{\textrm{F}}(p_x\sigma_x+p_y\sigma_y)$, where $p$ is momentum and $\sigma$ are Pauli matrices. It is obvious that a mass term $m\sigma_z$ will generate an energy gap for the electronic structure and the Dirac nodes are protected by extra physical symmetries apart from the lattice translational symmetry.

The search for 3D Dirac semimetal with simple electronic structure continues after the discovery of the 2D Dirac semimetals which can exist with or without spin-orbit coupling. The most direct generalization of 2D to 3D is the Weyl point of two bands, such as $H = v_{\textrm{F}}(p_x\sigma_x+p_y\sigma_y+p_z\sigma_z)$. The three Pauli matrices are all now used up in above equation and there is no local mass term. The Weyl semimetal phase is robust against perturbation and the robustness results from a topological (spin-orbit and symmetry together) consideration, which means that there cannot be only one Weyl point on the Fermi surface as the total Chern number must be zero per Brillouin zone (BZ). In materials with both time-reversal (T) and space inversion (I) symmetries, Weyl points must come together in pairs, degenerate in energy to form 3D Dirac points. Due to the inevitable degeneracy in T and I symmetric systems, one can only see 3D Dirac states rather than Weyl states. In order to achieve the Weyl states, one has to break either T or I symmetry.

Three-dimensional (3D) Dirac fermion metals, sometimes noted as the bulk Dirac semimetal phases, are of great interest if the material possesses 3D isotropic or anisotropic relativistic dispersion in the presence of strong spin-orbit coupling. It has been theoretically predicted that a topological (spin-orbit) 3D spin-orbit Dirac semimetal can be viewed as a composite of two sets of Weyl fermions where broken time-reversal or space inversion symmetry can lead to a surface Fermi-arc semimetal phase or a topological insulator \cite{Dai}. In the absence of spin-orbit coupling, topological phases cannot be derived from a 3D Dirac semimetal. Thus the parent bulk Dirac semimetal phase with strong spin-orbit coupling is of great interest. Moreover, it is theoretically predicted that Weyl semimetal (WS) phase can be existed in HgCr$_2$Se$_4$ \cite{Dai_WS}, Pyrochlore Iridates \cite{Japan, Wan} and $\beta$ -cristobalite BiO$_2$ \cite{Rappe}. Despite their predicted existence of bulk Dirac semimetal phase \cite{Murakami, Dirac_semi, Dai} and WS  phase \cite{Dai_WS, Japan, Wan, Rappe}, experimental studies have been lacking since it has been difficult to realize these phases in real materials, especially in stoichiometric single crystalline non-metastable systems with high mobility. It has also been noted that the bulk Dirac semimetal state can be achieved at the critical point of a topological phase transition \cite{Xu, Ando QPT, Oh} between a normal insulator and a topological insulator (Fig.~\ref{CdAs_1}), which requires fine-tuning of the chemical doping composition thus by effectively varying the spin-orbit coupling strength. This approach also introduces chemical disorder into the system. In stoichiometric bulk materials, the known 3D Dirac fermions in bismuth are in fact of massive variety since there clearly exists a band gap in the bulk Dirac spectrum \cite{ Bismuth}. On the other hand, the bulk Dirac fermions in the Bi$_{1-x}$Sb$_x$ system coexist with additional Fermi surfaces \cite{Hsieh1}.

Recently, several theoretical studies have predicted the existence of the topological Dirac semimetal (TDS) \cite{Dirac_3D, Dirac_semi, Dai}. In a 3D TDS, the 3D Dirac band touchings arise from the protection of certain space group crystalline symmetries, and are therefore proposed to be more robust to disorders or chemical alloying \cite{Dirac_3D, Dirac_semi, Dai, Xu, Oh}. Moreover, the topological Dirac semimetal differs from other types of 3D Dirac semimetals because it possesses strong spin-orbit coupling that leads to an inverted bulk band structure, making it possible to realize 3D Dirac multiplet states and host nontrivial topological order as well as novel spin-momentum locked Fermi arc surface states \cite{Dirac_semi, Dai}. In Ref. \cite{CdAs_Hasan}, we report experimental discovery of the gapless TDS phase in high-mobility stoichiometric material Cd$_3$As$_2$. Similar experimental results are also reported in Ref. \cite{CdAs_Cava}. Furthermore, experimental realization of the 3D Dirac phase in a metastable low mobility compound, Na$_3$Bi has also been reported \cite{Chen_Na3Bi, Hasan N3Bi} .

Fig.~\ref{CdAs_1}\textbf{a,b} show the crystal structure of Cd$_3$As$_2$, which has a tetragonal unit cell with $a= 12.67$ $\AA$ and $c= 25.48$ $\AA$ for $Z= 32$ with symmetry of space group $I4_1$cd. In this structure, arsenic ions are approximately cubic close-packed and Cd ions are tetrahedrally coordinated, which can be described in parallel to a fluorite structure of systematic Cd/As vacancies. There are four layers per unit and the missing Cd-As$_4$ tetrahedra are arranged without the central symmetry as shown with the (001) projection view in Fig.~\ref{CdAs_1}\textbf{b}, with the two vacant sites being at diagonally opposite corners of a cube face \cite{crys_str}.

In order to resolve a low-lying small dispersion feature near the Fermi level, we perform high-resolution ARPES dispersion measurements in the close vicinity of the Fermi level as shown in Fig.~\ref{CdAs_1}\textbf{c}. Remarkably, a linearly dispersive upper Dirac cone is observed at the surface BZ center $\bar{\Gamma}$ point, whose Dirac node is found to locate at a binding energy of $E_{\textrm{B}}\simeq0.2$ eV. At the Fermi level, only the upper Dirac band but no other electronic states are observed. On the other hand, the linearly dispersive lower Dirac cone is found to coexist with another parabolic bulk valence bands (Fig.~\ref{CdAs_1}\textbf{c} and Fig.~\ref{CdAs_2}\textbf{b}). From the observed steep Dirac dispersion (Fig.~\ref{CdAs_1}\textbf{c}), we obtain a surprisingly high Fermi velocity of about 9.8 eV$\cdot$$\AA$ ($\simeq1.5\times10^{6}$ ms$^{-1}$). This is more than 10-fold larger than the theoretical prediction of 0.15 eV$\cdot$ $\AA$  at the corresponding location of the chemical potential \cite{Dai}. Compared to the much-studied 2D Dirac systems, the Fermi velocity of the 3D Dirac fermions in Cd$_3$As$_2$ is thus about 3 times higher than that of in the topological surface states (TSS) of Bi$_2$Se$_3$ \cite{Xia}, $1.5$ times higher than in graphene \cite{Eli} and 30 times higher than that in the topological Kondo insulator phase in SmB$_6$ \cite{Dai_SmB6, MN}. The observed large Fermi velocity of the 3D Dirac band provides clues to understand Cd$_3$As$_2$'s unusually high mobility reported in previous transport experiments \cite{Mobi, Mobi2}.

In theory, there are two 3D Dirac nodes that are expected at two special $\mathbf{k}$ points along the $\Gamma-Z$ momentum space cut-direction (Figs.~\ref{CdAs_1}\textbf{d,e}). At the (001) surface, these two $\mathbf{k}$ points along the $\Gamma-Z$ axis project on to the $\bar{\Gamma}$ point of the (001) surface BZ (Fig.~\ref{CdAs_1}\textbf{d}). Therefore, at the (001) surface, theory predicts one 3D Dirac cone at the BZ center $\bar{\Gamma}$ point (Fig.~\ref{CdAs_2}\textbf{a}). These results are in qualitative agreement with our data, which supports our experimental observation of the 3D TDS phase in Cd$_3$As$_2$. We also study the ARPES measured constant energy contour maps (Figs.~\ref{CdAs_2}\textbf{c,d}). At the Fermi level, the constant energy contour consists of a single pocket centered at the $\bar{\Gamma}$ point. With increasing binding energy, the size of the pocket decreases and eventually shrinks to a point (the 3D Dirac point) near $E_{\textrm{B}}\simeq0.2$ eV. The observed anisotropies in the iso-energetic contours are likely due to matrix element effects associated with the standard p-polarization geometry used in our measurements.

The distinct semimetal nature of Cd$_3$As$_2$ is better understood from ARPES data if we compare our results with that of the prototype TI, Bi$_2$Se$_3$. In Bi$_2$Se$_3$ as shown in Fig.~\ref{CdAs_3}\textbf{b}, the bulk conduction and valence bands are fully separated (gapped), and a linearly dispersive topological surface state is observed that connect across the bulk band-gap. In the case of Cd$_3$As$_2$ (Fig.~\ref{CdAs_3}\textbf{a}), there does not exist a full bulk energy gap. On the other hand, the bulk conduction and valence bands ``touch'' (and only ``touch'') at one specific location in the momentum space, which is the 3D band-touching node, thus realizing a 3D TDS. For comparison, we further show that a similar TDS state is also realized by tuning the chemical composition $\delta$ (effectively the spin-orbit coupling strength) to the critical point of a topological phase transition between a normal insulator and a topological insulator. Fig.~\ref{CdAs_3}\textbf{c,d} present the surface electronic structure of two other TDS phases in the BiTl(S$_{1-\delta}$Se$_{\delta}$)$_2$ and (Bi$_{1-\delta}$In$_{\delta}$)$_2$Se$_3$ systems. In both systems, it has been shown that tuning the chemical composition $\delta$ can drive the system from a normal insulator state to a topological insulator state \cite{Xu, Ando QPT, Oh}. The critical compositions for the two topological phase transitions are approximately near $\delta=0.5$ and $\delta=0.04$, respectively. Fig.~\ref{CdAs_3}\textbf{c,d} show the ARPES measured surface electronic structure of the critical compositions for both BiTl(S$_{1-\delta}$Se$_{\delta}$)$_2$ and (Bi$_{1-\delta}$In$_{\delta}$)$_2$Se$_3$ systems, which are expected to exhibit the TDS phase. Indeed, the bulk critical compositions where bulk and surface Dirac bands collapse also show Dirac cones with intensities filled inside the cones, which is qualitatively similar to the case in Cd$_3$As$_2$.

Based on the ARPES data in Fig.~\ref{CdAs_2}\textbf{c,d}, the Fermi velocity is estimated to be $\sim4$ eV$\cdot\textrm{\AA}$ and $\sim2$ eV$\cdot\textrm{\AA}$ for the 3D Dirac fermions in BiTl(S$_{1-\delta}$Se$_{\delta}$)$_2$ and (Bi$_{1-\delta}$In$_{\delta}$)$_2$Se$_3$ respectively, which is much lower than that of what we observe in Cd$_3$As$_2$, thus likely limiting the carrier mobility. The mobility is also limited by the disorder due to strong chemical alloying. More importantly, the fine control of doping/alloying $\delta$ value and keeping the composition exactly at the bulk critical composition is difficult to achieve \cite{Xu}, especially while considering the chemical inhomogeneity introduced by the dopants.
We have experimentally identified the crystalline-symmetry-protected 3D spin-orbit TDS phase in a stoichiometric system Cd$_3$As$_2$. Our experimental identification of the Dirac-like bulk topological semimetal phase in high mobility Cd$_2$As$_2$ opens the door for exploring higher dimensional spin-orbit Dirac physics in a real material

\section{Topological Crystalline Insulators}

In this section, we review the research on the topological crystalline insulator (TCI). In particular, we focus on the mirror symmetry protected TCI phase and its experimental discovery in the Pb$_{1-x}$Sn$_x$Te(Se) system. We review the methodology developed to uniquely determine the topological number (the mirror Chern number $n_M$) in Pb$_{1-x}$Sn$_x$Te by measuring its surface state spin texture including the chirality (or handness) using spin-resolved ARPES.

The 3D Z$_2$ (Kane-Mele) topological insulator represents the first example in nature of a topologically ordered electronic phase existing in bulk solids \cite{RMP}. In a 3D  Z$_2$ (Kane-Mele) TI, it is the protection of time-reversal symmetry that gives rise to a nontrivial Z$_2$ topological invariant. With the explosion of research interest on 3D Z$_2$ TI materials, a new research topic that focuses on searching for new topologically nontrivial phases protected by other discrete symmetries emerged. In 2011, a new topological phase of matter, which is now usually referred as the topological crystalline insulator (TCI), was theoretically proposed by Fu \cite{Liang PRL TCI}. In a TCI, space group symmetries of the crystalline system replace the role of time-reversal symmetry in an otherwise Z$_2$ TI. Therefore, the TCI phase is topologically distinct from the much-studied Z$_2$ TI, and it is believed to host many exotic topological quantum properties, such as higher order (non-linear) surface band crossings, topological state without spin-orbit coupling, and crystalline symmetry protected topological superconductivity or Chern currents \cite{Liang PRL TCI}. So far, possible TCI phases have been theoretically discussed for systems possessing four-fold ($C_4$) or six-fold ($C_6$) rotational symmetry as well as the mirror symmetry \cite{Liang PRL TCI, Liang NC SnTe}. And the mirror symmetry case gained particular interests since a real material prediction, namely the Pb$_{1-x}$Sn$_x$Te system, was made by Hsieh \textit{et al} via first-principles band structure calculations \cite{Liang NC SnTe}.

\begin{figure*}
\centering
\includegraphics[width=17cm]{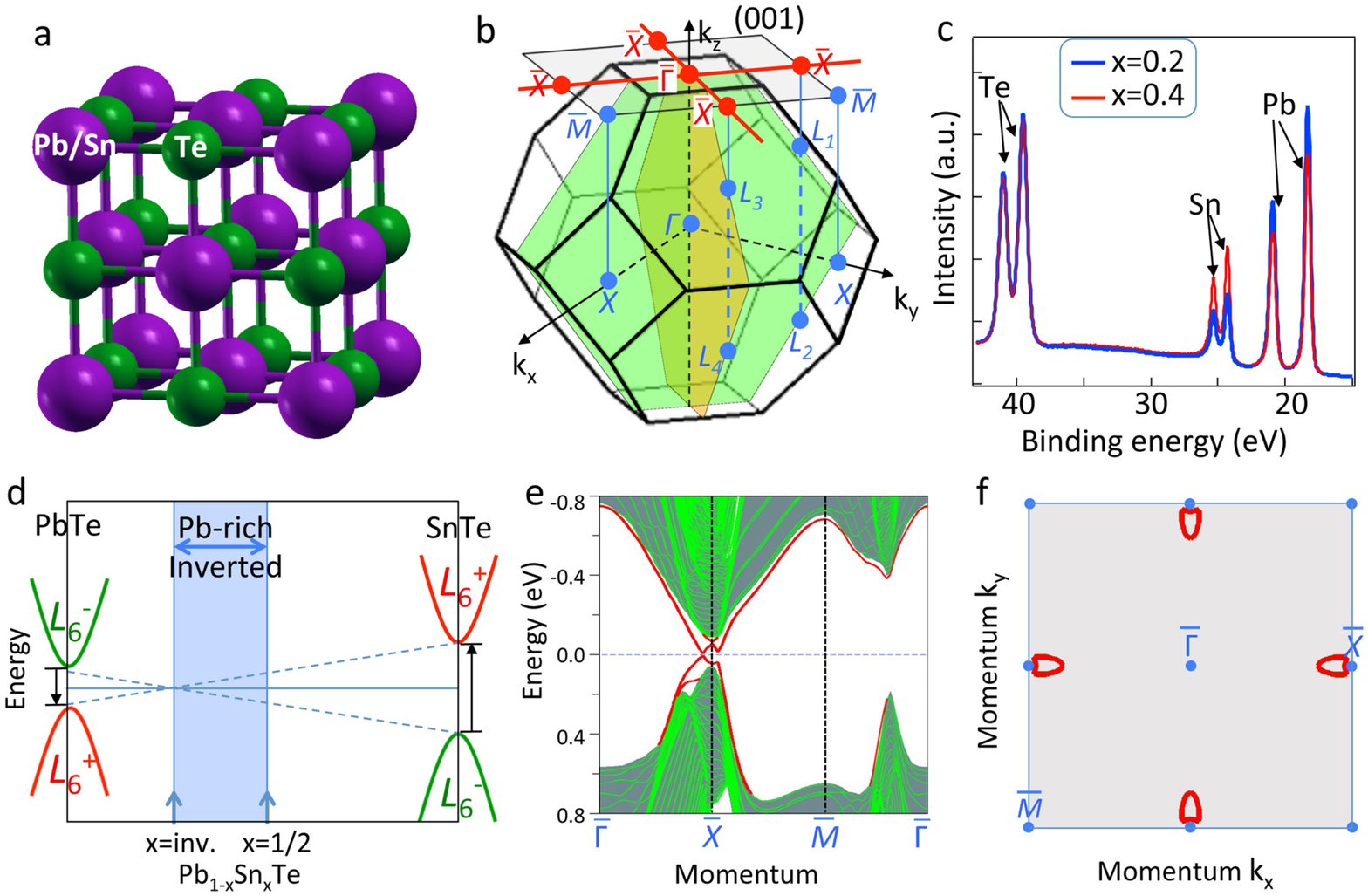}
\caption{\label{FCC} \textbf{Band inversion transition and double Dirac surface states in Pb$_{1-x}$Sn$_{x}$Te.} \textbf{a,} The lattice of Pb$_{1-x}$Sn$_{x}$Te system is based on the ``sodium chloride'' crystal structure. The Pb-rich side of the Pb$_{1-x}$Sn$_{x}$Te possesses the ideal ``sodium chloride'' crystal structure without rhombohedral distortion. \textbf{b,} The first Brillouin zone (BZ) of Pb$_{1-x}$Sn$_{x}$Te lattice. The mirror planes are shown using green and light-brown colors. These mirror planes project onto the (001) crystal surface as the $\bar{\textrm{X}}-\bar{\Gamma}-\bar{\textrm{X}}$ mirror lines. \textbf{c,} ARPES measured core level spectra (incident photon energy 75 eV) of two representative compositions, namely Pb$_{0.8}$Sn$_{0.2}$Te and Pb$_{0.6}$Sn$_{0.4}$Te. The photoemission (spin-orbit coupled) core levels of tellurium 4d, tin 4d, and lead 5d orbitals are observed. \textbf{d,} The bulk band-gap of Pb$_{1-x}$Sn$_{x}$Te alloy system undergoes a band inversion upon changing the Pb/Sn ratio. A TCI phase with metallic surface states is theoretically predicted when the band-gap is inverted (toward SnTe) \cite{Liang NC SnTe}. The Pb-rich inverted regime lies on the inverted compositional range yet still with Pb\%$>$Sn\% ($x_{\textrm{inversion}}<x<1/2$). \textbf{e,f,} First-principles based calculation of band dispersion \textbf{(e)} and iso-energetic contour with energy set 0.02eV below the Dirac node energy \textbf{(f)} of the inverted end compound SnTe as a qualitative reference for the ARPES experiments. The surface states are shown by the red lines whereas the bulk band projections are represented by the green shaded area in \textbf{e}. Adapted from S.-Y. Xu $et$ $al.$, \textit{Nature Commun.} $\mathbf{3}$, 1192 (2012) \cite{TCI Hasan}.}
\end{figure*}

Pb$_{1-x}$Sn$_x$Te is a pseudobinary semiconducting system widely used for infrared optoelectronic and thermoelectric devices. It is known that the band-gap at the four L points in the bulk Brillouin zone (BZ) closes itself and re-opens upon increasing $x$ in the Pb$_{1-x}$Sn$_{x}$Te system \cite{PST Inversion1} (Fig.~\ref{FCC}). The fact that band inversion occurs at even number of points per bulk BZ excludes the possibility of the Z$_2$-type (Kane-Mele) topological insulator phase in the Pb$_{1-x}$Sn$_{x}$Te system under ambient pressure \cite{RMP}. However, Hsieh \textit{et al} noticed that any two of the four L points along with the $\Gamma$ point form a momentum-space mirror plane, making it possible to realize a novel topological phase related to the crystalline mirror symmetry in Pb$_{1-x}$Sn$_{x}$Te \cite{Liang NC SnTe}. Detailed theoretical analysis in Ref. \cite{Liang NC SnTe} showed that the Pb$_{1-x}$Sn$_{x}$Te system can theoretically host a unique mirror symmetry protected TCI phase with a nontrivial topological invariant that is the mirror Chern number $n_M$, whereas the Z$_2$ invariant $\nu_0$ equals to $0$ for Pb$_{1-x}$Sn$_x$Te showing the predicted TCI phase's irrelevance to time-reversal symmetry. Therefore, the experimental identification of the mirror symmetry protected TCI phase in Pb$_{1-x}$Sn$_x$Te requires to not only observe surface states within the bulk energy gap but also find a way that can uniquely measure its topological number $n_M$.

It turns out that studying the surface state spin polarization and its momentum-space texture chirality (or handness) serves as keys to probing the role of topology in the predicted TCI phase in Pb$_{1-x}$Sn$_{x}$Te. This is because of the distinct property of its topological number, namely the mirror Chern number $n_M$. Unlike the Z$_2$ invariant $\nu_0$ that can only be 0 or 1, the mirror Chern number $n_M$ can take any integer value. While the absolute value of $n_M$ is determined by the number of surface states that disperse along each momentum-space mirror direction, the sign of $n_M$ is uniquely fixed by the chirality of the surface state spin texture \cite{Liang NC SnTe, Science, 20}. Furthermore, because of the predicted four band inversions in Pb$_{1-x}$Sn$_{x}$Te, it is in principle more favorable to study the inverted end-compound SnTe because it has the largest inverted band-gap. However, it has been known that SnTe is heavily $p-$type, due to the fact that Sn vacancies are thermodynamically stable \cite{SnTe p-type}, which makes the chemical potential cut deeply inside the bulk valence bands \cite{SnTe ARPES} (see Fig.~\ref{FCC}). Therefore, one needs to work with the system in the Pb-rich (yet still inverted) regime, in order to access and study the predicted surface states via photoemission experiments.

Following the theoretical prediction in Ref. \cite{Liang NC SnTe}, ARPES experiments have been performed in Pb$_{1-x}$Sn$_{x}$Te, Pb$_{1-x}$Sn$_{x}$Se and SnTe \cite{TCI Hasan, TCI Story, TCI Ando}, and the existence of Dirac surface states inside the bulk energy gap has been observed in both Pb-rich Pb$_{0.6}$Sn$_{0.4}$Te \cite{TCI Hasan} and Pb$_{0.77}$Sn$_{0.23}$Se \cite{TCI Story} systems. More importantly, the helical spin texture and its chirality (or handness) have been systematically mapped out by spin-resolved ARPES experiments in Pb$_{0.6}$Sn$_{0.4}$Te.

\begin{figure*}
\centering
\includegraphics[width=16cm]{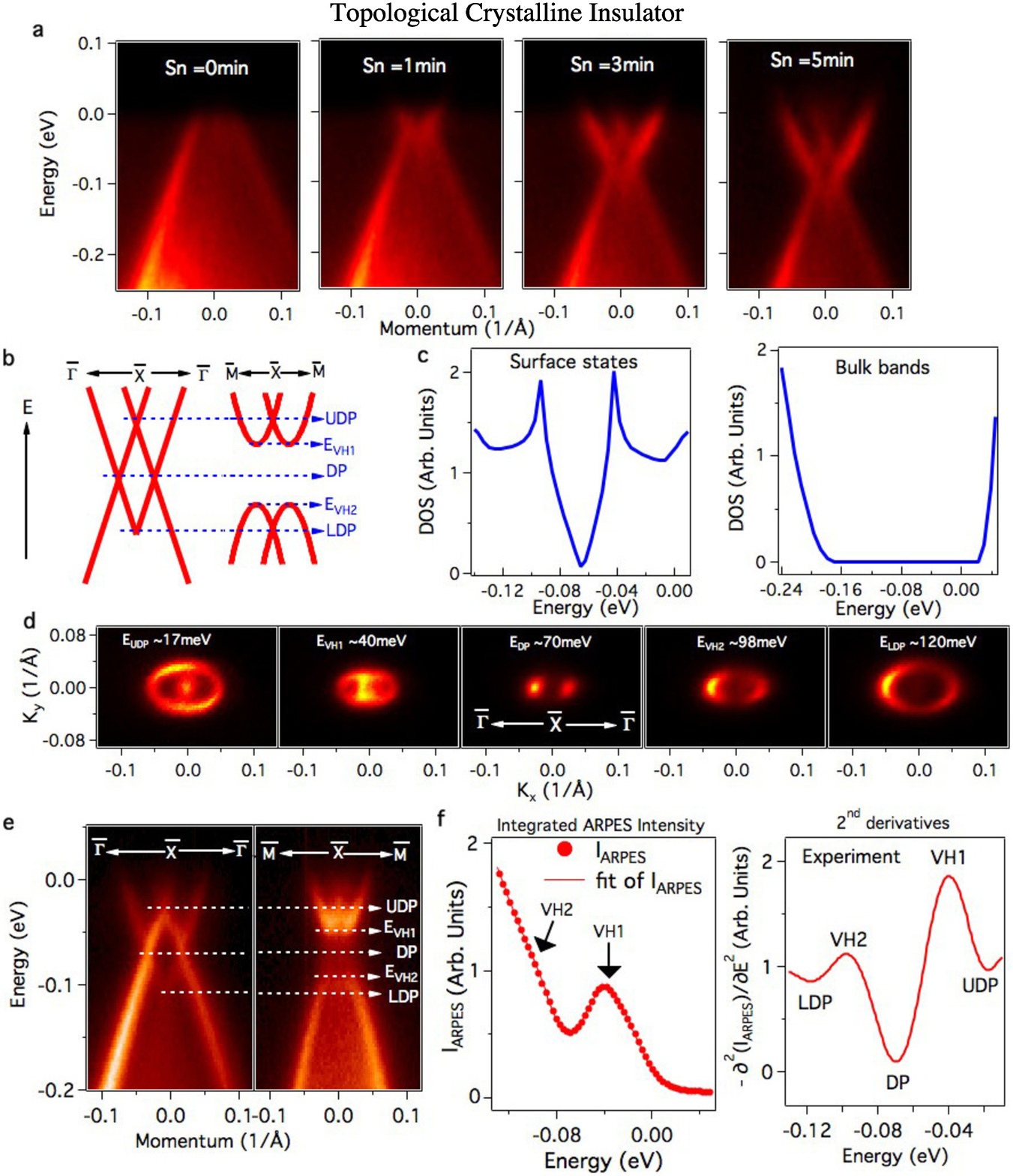}
\caption{\label{Saddle}\textbf{Observation of the topological crystalline surface states and saddle point singularity.} \textbf{a,} ARPES dispersion maps upon \textit{in situ} Sn deposition on the Pb$_{0.70}$Sn$_{0.30}$Se surface. The dosage (time) for Sn deposition is noted. A different batch of sample, which is $p-$type with the chemical potential below the Dirac points, is used for the Sn deposition data shown in this panel. \textbf{b and c,} Schematics of surface band dispersion of the TCI phase along the mirror line $\bar\Gamma-\bar{\textrm{X}}-\bar\Gamma$ and the $\bar{\textrm{M}}-\bar{\textrm{X}}-\bar{\textrm{M}}$ momentum space cut-directions. Five important features of the surface states, including Dirac point of the upper part of the Dirac cones (UDP), van Hove singularity of the upper Dirac cones (VH1), two Dirac points along the $\bar\Gamma-\bar{\textrm{X}}-\bar\Gamma$ mirror line (DP), van Hove singularity of the lower part of the Dirac cones (VH2) and Dirac point of the lower part of the Dirac cones (LDP) are marked. \textbf{c,} Calculated density of state (DOS) for the surface states and the bulk bands using the $k\cdot p$ model. }
\end{figure*}

Utilizing spin-resolved angle-resolved photoemission spectroscopy, Xu \textit{et al} for the first time experimentally determined the topological mirror Chern number of $n_M=-2$ in Pb$_{1-x}$Sn$_{x}$Te(Se), which experimentally revealed its topological mirror nontriviality of the TCI phase in the Pb$_{1-x}$Sn$_{x}$Te(Se) system. The experimental data were reported in Ref. \cite{TCI Hasan} and are summarized in Figs.~\ref{FCC}-\ref{TCITI}. As shown in Fig.~\ref{Saddle}, two distinct surface states that cross the Fermi level are observed on the opposite sides of each $\bar{X}$ point along the $\bar{\Gamma}-\bar{X}-\bar{\Gamma}$ direction at the (001) surface of Pb$_{0.7}$Sn$_{0.3}$Se. Therefore, in total four surface states are observed within one surface BZ, consistent with the predicted four bulk band inversions. All surface states are observed to locate along the momentum space mirror line direction $\bar{\Gamma}-\bar{X}-\bar{\Gamma}$, which reflects the mirror symmetry protection to the observed topological surface states. Each $\bar{\Gamma}-\bar{X}-\bar{\Gamma}$ mirror line possesses two surface states, from which the absolute value of the topological mirror Chern number of $|n_M|=2$ is determined. It is also interesting to notice that since there are two surface Dirac cones that are located very close to near each $\bar{X}$ point, they inevitably touch and hybridize with each other, giving rise to a topological change in the band contours, also known as a Lifshitz transition in the electronic structure. Such Fermi surface Lifshitz transition is clearly observed in Fig.~\ref{Saddle}\textbf{d}. At the energy where the Lifshitz transition happens, a saddle point type of van Hove singularity (VHS) is expected leading to the divergence of the density of states at the energy. Such VHS is also observed in our ARPES data shown in Fig.~\ref{Saddle}\textbf{f}. Observation of saddle point singularity on the surface of Pb$_{0.7}$Sn$_{0.3}$Se paves the way for realizing correlated physical phenomena in topological Dirac surface states.

To uniquely determine the topological mirror Chern number $n_M$, spin-resolved ARPES measurements were performed on the TCI surface states in Pb$_{0.6}$Sn$_{0.4}$Te as shown in Fig.~\ref{TCITI}. Fig.~\ref{TCITI}\textbf{h} shows that four distinct spin polarizations with the configuration of $\downarrow, \uparrow, \downarrow, \uparrow$ are observed along the mirror line $\bar{\Gamma}-\bar{X}-\bar{\Gamma}$. These measurements clearly identify the one to one helical spin-momentum locking in the surface states. More importantly, the right-handed chirality of the spin texture is experimentally measured for the lower-Dirac-cone states, which therefore determines the topological mirror Chern number of $n_M=-2$. These systematic measurements reported in Ref. \cite{TCI Hasan} for the first time conclusively identified a novel mirror symmetry protected TCI phase by measuring its topological mirror number using a spin-sensitive probe.

We present a comparison of the Pb$_{0.6}$Sn$_{0.4}$Te and a single Dirac cone Z$_2$ topological insulator (TI) system GeBi$_2$Te$_4$ \cite{Ternary arXiv, Ternary PRB}. As shown in Fig.~\ref{TCITI}\textbf{a-d}, for the Z$_2$ TI system GeBi$_2$Te$_4$, a single surface Dirac cone is observed enclosing the time-reversal invariant Kramers' momenta $\bar{\Gamma}$ in both ARPES and calculation results, demonstrating its Z$_2$ topological insulator state and the time-reversal symmetry protection of its single Dirac cone surface states. On the other hand, for the Pb$_{0.6}$Sn$_{0.4}$Te samples (Fig.~\ref{TCITI}\textbf{e-h}), none of the surface states is observed to enclose any of the time-reversal invariant momentum, suggesting their irrelevance to the time-reversal symmetry related protection. With future ultra-high-resolution experimental studies to prove the strict gapless nature of the Pb$_{0.6}$Sn$_{0.4}$Te surface states and therefore their predicted topological protection by the crystalline mirror symmetries, it is then possible to realize magnetic yet topologically protected surface states in the Pb$_{1-x}$Sn$_{x}$Te system due to its irrelevance to the time-reversal symmetry related protection, which is fundamentally not possible in the Z$_2$ topological insulator systems.

The experimental discovery of mirror-protected TCI phase in the Pb$_{1-x}$Sn$_{x}$Te(Se) systems have attracted much interest in condensed matter physics and opened the door for many further theoretical and experimental studies on this novel TCI phase \cite{Vidya-1, Chen QPI, Liang PRB, Ali, Ong, SnTe Spin, Valla TCI, Saddle, PSS Spin, PSS Spin-2, Vidya-2, SnTe111-1, SnTe111-2, PSS111, In-SnTe-0, In-SnTe-1, In-SnTe-2, In-SnTe-3, In-SnTe-4, Chen, Spin-filter}. These following works include scanning tunneling spectroscopies \cite{Vidya-1, Chen QPI, Liang PRB, Ali}, thermal and electrical transport \cite{Ong}, further systematic studies on the surface spin and orbital textures \cite{SnTe Spin, PSS Spin, PSS Spin-2, Vidya-2}, as well as theoretical and experimental efforts in realizing mirror symmetry protected topological superconductivity \cite{In-SnTe-1, In-SnTe-2, In-SnTe-3, In-SnTe-4} or magnetic Chern current on the TCI surfaces \cite{Chen, Spin-filter}.

\addtocounter{figure}{-1}
\begin{figure*} [t!]
\caption{(Previous page.) \textbf{d,} Experimental observation of the Lifshitz transition - the binding energies are noted on the constant energy contours. \textbf{e,} ARPES measured dispersion plots along $\bar\Gamma-\bar{\textrm{X}}-\bar\Gamma$ and $\bar{\textrm{M}}-\bar{\textrm{X}}-\bar{\textrm{M}}$.  \textbf{f,} Momentum ($k_x$ and $k_y$) integrated ARPES intensity as a function of binding energy (left). 2$^{nd}$ derivative of the ARPES intensity with respect to binding energy is presented to further highlight the features. The upper Dirac point (UDP), upper van Hove singularity (VH1), Dirac point (DP), lower van Hove singularity (VH2) and lower Dirac point (LDP) are marked. Adapted from M. Neupane $et$ $al.$, Preprint at http://arXiv:1403.1560 (2014) \cite{Saddle}.}
\end{figure*}

\begin{figure*}
\includegraphics[width=17cm]{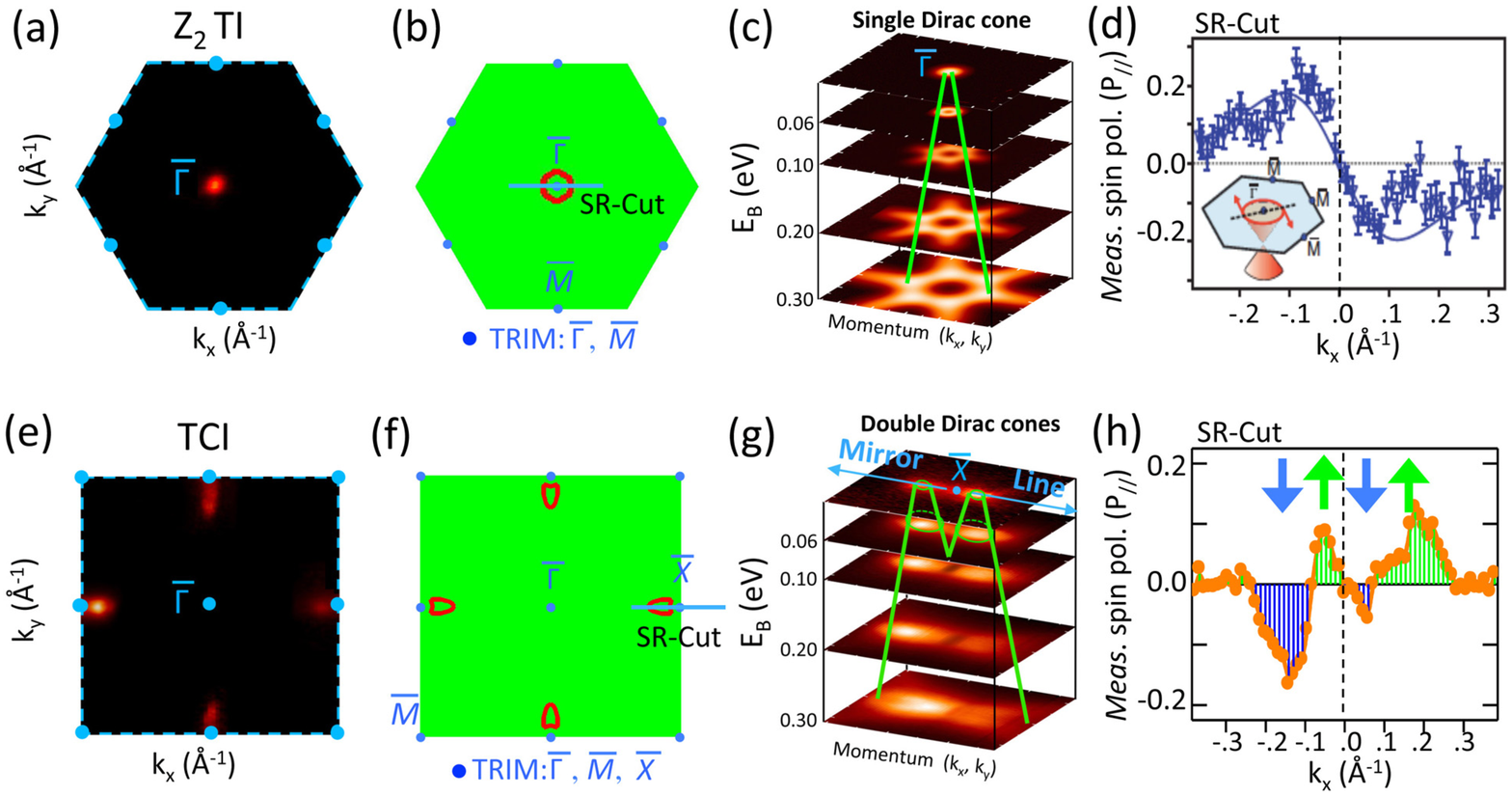}
\caption{\label{TCITI}\textbf{The topological distinction between Z$_2$ (Kane-Mele) topological insulator and topological crystalline insulator phases.} (a-d) ARPES, spin-resolved ARPES and calculation results of the surface states of a Z$_2$ topological insulator GeBi$_2$Te$_4$ \cite{Ternary arXiv}, an analog to Bi$_2$Se$_3$ \cite{Xia}. (a) ARPES measured Fermi surface with the chemical potential tuned near the surface Dirac point. (b) First-principles calculated iso-energetic contour of the surface states near the Dirac point. The solid blue line shows the momentum-space cut used for spin-resolved measurements. Right: A stack of ARPES iso-energetic contours near the $\bar{\Gamma}$ point of the surface BZ. (d) Measured spin polarization of Bi$_2$Se$_3$, in which a helical spin texture is revealed. (e-h)  ARPES and spin-resolved ARPES measurements on the Pb$_{0.6}$Sn$_{0.4}$Te ($x=0.4$) samples and band calculation results on the end compound SnTe \cite{Liang NC SnTe}. (e) ARPES measured Fermi surface map of Pb$_{0.6}$Sn$_{0.4}$Te. (f) First-principles calculated iso-energetic contour of SnTe surface states near the Dirac point. The solid blue line shows the momentum-space cut near the surface BZ edge center $\bar{X}$ point, which is used for spin-resolved measurements shown in panel (h). (g) A stack of ARPES iso-energetic contours near the $\bar{X}$ point of the surface BZ, revealing the double Dirac cone contours near each $\bar{X}$ point on the surface of Pb$_{0.6}$Sn$_{0.4}$Te. (h) Measured spin polarization of Pb$_{0.6}$Sn$_{0.4}$Te near the native Fermi energy along the momentum space cut defined in panel (f), in which two spin helical  Dirac cones are observed near an $\bar{X}$ point. [Adapted from S.-Y. Xu $et$ $al.$, \textit{Nature Commun.} \textbf{3}, 1192 (2012). \cite{TCI Hasan}]}
\end{figure*}

\section{Magnetic and Superconducting doped Topological Insulators}

\begin{figure*}
\includegraphics[width=17cm]{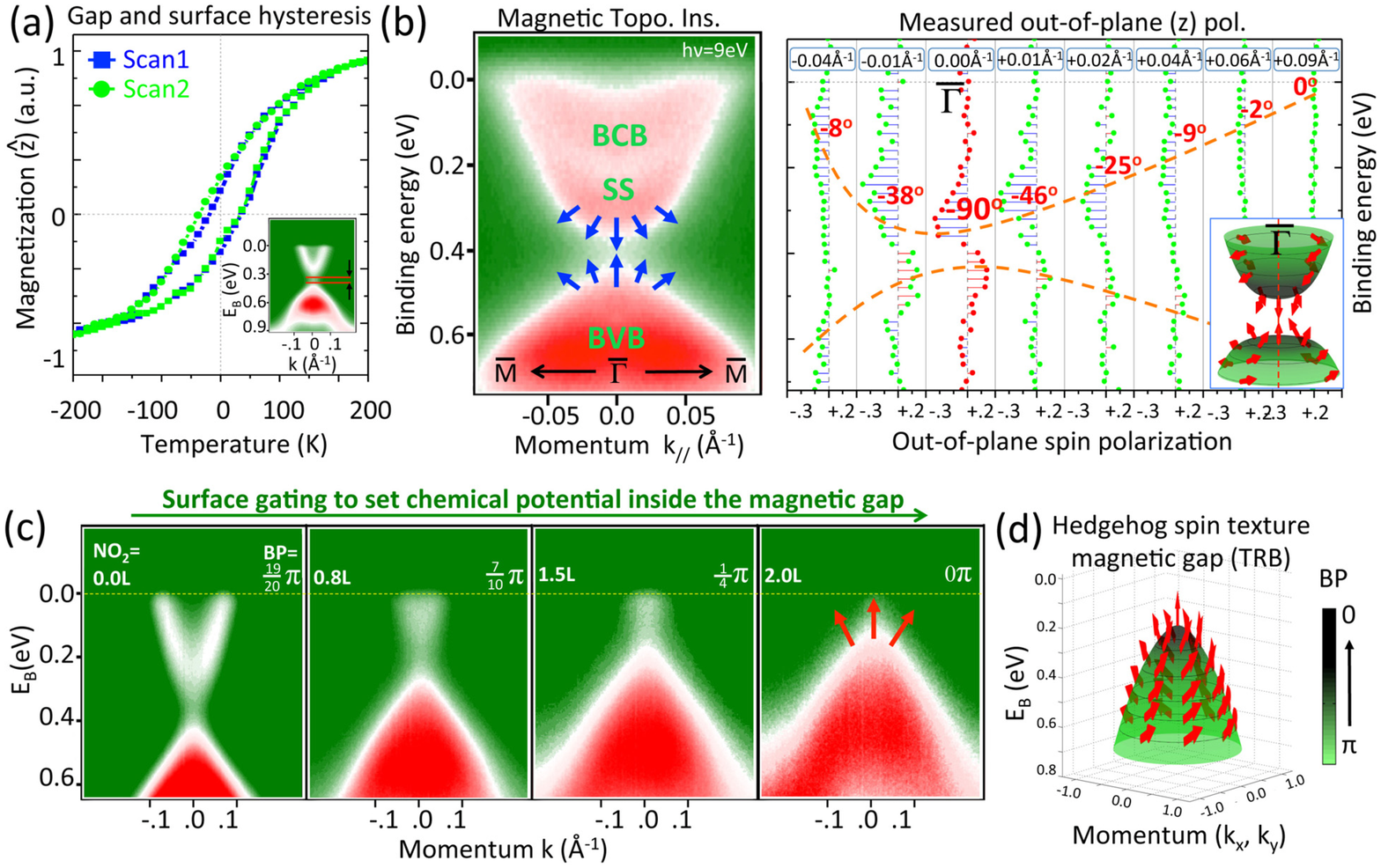}
\caption{\label{Hedgehog}\textbf{Hedgehog spin texture and Berry's phase tuning in a magnetic topological insulator.} (a) Magnetization measurements using magnetic circular dichroism shows out-of-plane ferromagnetic character of the Mn-Bi$_2$Se$_3$ MBE film surface through the observed hysteretic response. The inset shows the ARPES observed gap at the Dirac point in the Mn(2.5\%)-Bi$_2$Se$_3$ film sample. (b) Spin-integrated and spin-resolved measurements on a representative piece of Mn(2.5\%)-Bi$_2$Se$_3$ film sample using 9 eV photons. Left: Spin-integrated ARPES dispersion map. The blue arrows represent the spin texture configuration in close vicinity of the gap revealed by our spin-resolved measurements. Right, Measured out-of-plane spin polarization as a function of binding energy at different momentum values. The momentum value of each spin polarization curve is noted on the top. The polar angles ($\theta$) of the spin polarization vectors obtained from these measurements are also noted. The $90^{\circ}$ polar angle observed at $\bar{\Gamma}$ point suggests that the spin vector at $\bar{\Gamma}$ is along the vertical direction. The spin behavior at $\bar{\Gamma}$ and its surrounding momentum space reveals a hedgehog-like spin configuration for each Dirac band separated by the gap. Inset shows a schematic of the revealed hedgehog-like spin texture. (c) Measured surface state dispersion upon \textit{in situ} NO$_2$ surface adsorption on the Mn-Bi$_2$Se$_3$ surface. The NO$_2$ dosage in the unit of Langmuir ($1\textrm{L}=1\times10^{-6}$ torr${\cdot}$sec) and the tunable Berry's phase (BP) associated with the topological surface state are noted on the top-left and top-right corners of the panels, respectively. The red arrows depict the time-reversal breaking out-of-plane spin texture at the gap edge based on the experimental data. (d) The time-reversal breaking spin texture features a singular hedgehog-like configuration when the chemical potential is tuned to lie within the magnetic gap, corresponding to the experimental condition presented in the last panel in panel (c). [Adapted from S.-Y. Xu $et$ $al.$, \textit{Nature Physics} \textbf{8}, 616 (2012). \cite{Hedgehog}].}
\end{figure*}

In this section, we review the photoemission studies on magnetic or superconducting topological insulators. The goal of the ARPES and spin-resolved ARPES studies on magnetic topological insulators is to resolve the magnetic gap opened at the surface Dirac point as well as the magnetically-driven spin texture near the gap edge. The magnetic gap and its spin texture are the keys to realizing the proposed novel effects based on a magnetic topological insulator, including quantum anomalous Hall effect \cite{Yu Science QAH} and topological magneto-electrical effect \cite{Essin PRL, Qi PRB, Zhang Axion}. On the other hand, the goal of ARPES studies on superconducting topological insulators is to resolve the superconducting gap in the topological surface states, which is predicted as a promising platform in realizing Majorana fermion modes \cite{Kane_Proximity, Zhang_TSC}.

We first focus on the research on magnetic topological insulators. Since the discovery of three dimensional topological insulators \cite{RMP}, topological order proximity to ferromagnetism or superconductivity has been considered as one of the core interest of the field \cite{Qi PRB, Yu Science QAH, Galvanic effect, Essin PRL, Zhang Axion, Hor PRB BiMnTe, Yayu WAL, Fe XMCD, Wray2, Checkelsky, Xue Science QAH}. Such interest is strongly motivated by the proposed time-reversal (TR) breaking topological physics such as quantized anomalous chiral Hall current, spin current, axion electrodynamics, and inverse spin-galvanic effect \cite{Qi PRB, Yu Science QAH, Galvanic effect, Essin PRL, Zhang Axion}, all of which critically rely on finding a way to break TR symmetry to open up a magnetic gap at the Dirac point on the surface and to further utilize the unique TR broken spin texture for applications.

Experimentally, a number of photoemission experiments have been performed in magnetically doped topological insulators, in order to observe the energy gap at the Dirac point opened by the breaking time-reversal symmetry via magnetic doping. Although gap-like feature at the Dirac point has been reported and interpreted as the magnetic gap \cite{Wray2, Ando QPT}, a number of other factors, such as spatial fluctuation of momentum and energy near the Dirac point \cite{Haim Nature physics BiSe} and surface chemical modifications \cite{Wray2, Yayu Science}, contribute to the observed gap \cite{Wray2, Haim Nature physics BiSe, vdW, Gap, Yayu Science, Hofmann}. The photoemission probe previously used to address the gap cannot distinguish or isolate these factors that respect TR symmetry from the TR breaking effect as highlighted in recent STM works \cite{Haim Nature physics BiSe}. In fact, photoemission Dirac point spectral suppression including a gap is also observed even on stoichiometric TI crystals without magnetic dopants or ferromagnetism \cite{Gap}. This is because surface can acquire nontrivial energy gaps due to ad-atom hybridization, surface top layer relaxation, Coulomb interaction from deposited atoms, and other forms of surface chemistry such as \textit{in situ} oxidation \cite{Wray2, Haim Nature physics BiSe, vdW, Gap, Yayu Science, Hofmann}. Under such conditions, it was not possible to isolate TR breaking effect from the rest of the extrinsic surface gap phenomena \cite{Wray2, Haim Nature physics BiSe, vdW, Gap, Yayu Science, Hofmann}. Therefore, the establishment of TR breaking effect fundamentally requires measurements of electronic groundstate with a \textit{spin}-sensitive probe.

In Ref. \cite{Hedgehog}, the authors utilized spin-resolved angle-resolved photoemission spectroscopy to measure the momentum space spin configurations in systematically magnetically doped, non-magnetically doped, and ultra-thin quantum coherent topological insulator films, in order to understand the nature of electronic groundstates under two extreme limits vital for magnetic topological devices. These measurements allow to make definitive conclusions regarding magnetism on topological surfaces, and make it possible to quantitatively \textit{isolate} the TR breaking effect in generating the surface electronic gap from many other physical or chemical changes also leading to gap-like behavior \cite{Ando QPT, Haim Nature physics BiSe, vdW, Hofmann} often observed on the surfaces. Spin reorientation measurements and the systematic methodology demonstrated here can be utilized to probe quantum magnetism on the surfaces of other materials as well. Furthermore, following this spin-resolved ARPES work \cite{Hedgehog}, surface magnetism mediated by the surface Dirac fermions were again confirmed by transport experiments \cite{Checkelsky}. And very recently, the long-sought quantum anomalous Hall currents have been observed in magnetically-doped topological insulator thin films \cite{Xue Science QAH}.

Fig.~\ref{Hedgehog} presents the key spin-resolved measurements on magnetically-doped topological insulator thin films, which reveals the exotic time-reversal breaking (hedgehog-like \cite{Hedgehog}) spin texture near the edge of the magnetic gap. Fig.~\ref{Hedgehog}\textbf{a} shows a hysteretic measurement using x-ray circular dichroism in the out-of-plane direction, which suggests a ferromagnetically ordered groundstate mediated by the surface Dirac fermions \cite{Checkelsky}. Fig.~\ref{Hedgehog}\textbf{b} shows the out-of-plane spin polarization ($P_z$) measurements of the electronic states in the vicinity of the Dirac point gap of a Mn(2.5\%)-Bi$_2$Se$_3$ sample. The surface electrons at the time-reversal invariant $\bar{\Gamma}$ point (red curve in Fig.~\ref{Hedgehog}\textbf{b}) are clearly observed to be spin polarized in the out-of-plane direction. The opposite sign of $P_z$ for the upper and lower Dirac band shows that the Dirac point spin degeneracy is indeed lifted up ($\textrm{E}(k_{//}=0,\uparrow){\neq}\textrm{E}(k_{//}=0,\downarrow)$), which manifestly breaks the time-reversal symmetry on the surface of our Mn(2.5\%)-Bi$_2$Se$_3$ samples. Systematic spin-resolved measurements as a function of binding energy and momentum reveal a Hedgehog-like spin texture (inset of Fig.~\ref{Hedgehog}\textbf{b}). As demonstrated recently \cite{Xu}, the quantum Berry's phase (BP) defined on the spin texture of the surface state Fermi surface bears a direct correspondence to the bulk topological invariant realized in the bulk electronic band structure via electronic band inversion \cite{Nature_2009, Xu}. We experimentally show that a BP tunability can be realized on our magnetic films which is important to prepare the sample condition to the axion electrodynamics limit. On the Mn-Bi$_2$Se$_3$ film, spin configuration pattern can be understood as a competition between the out-of-plane TR breaking component and the in-plane helical component of spin. The in-plane spin that can be thought of winding around the Fermi surface in a helical pattern contributes to a nonzero BP \cite{Nature_2009}, whereas the out-of-plane TR breaking spin direction is constant as one loops around the Fermi surface hence does not contribute to the Berry's phase (BP). Such exotic spin groundstate in a magnetic topological insulator enables a tunable Berry's phase on the magnetized topological surface \cite{Hedgehog}, as experimentally demonstrated by our chemical gating via NO$_2$ surface adsorption method shown in Figs.~\ref{Hedgehog}\textbf{c,d}.

\begin{figure*}
\includegraphics[scale=0.3,clip=true, viewport=0.0in 0.0in 20.5in 13.5in]{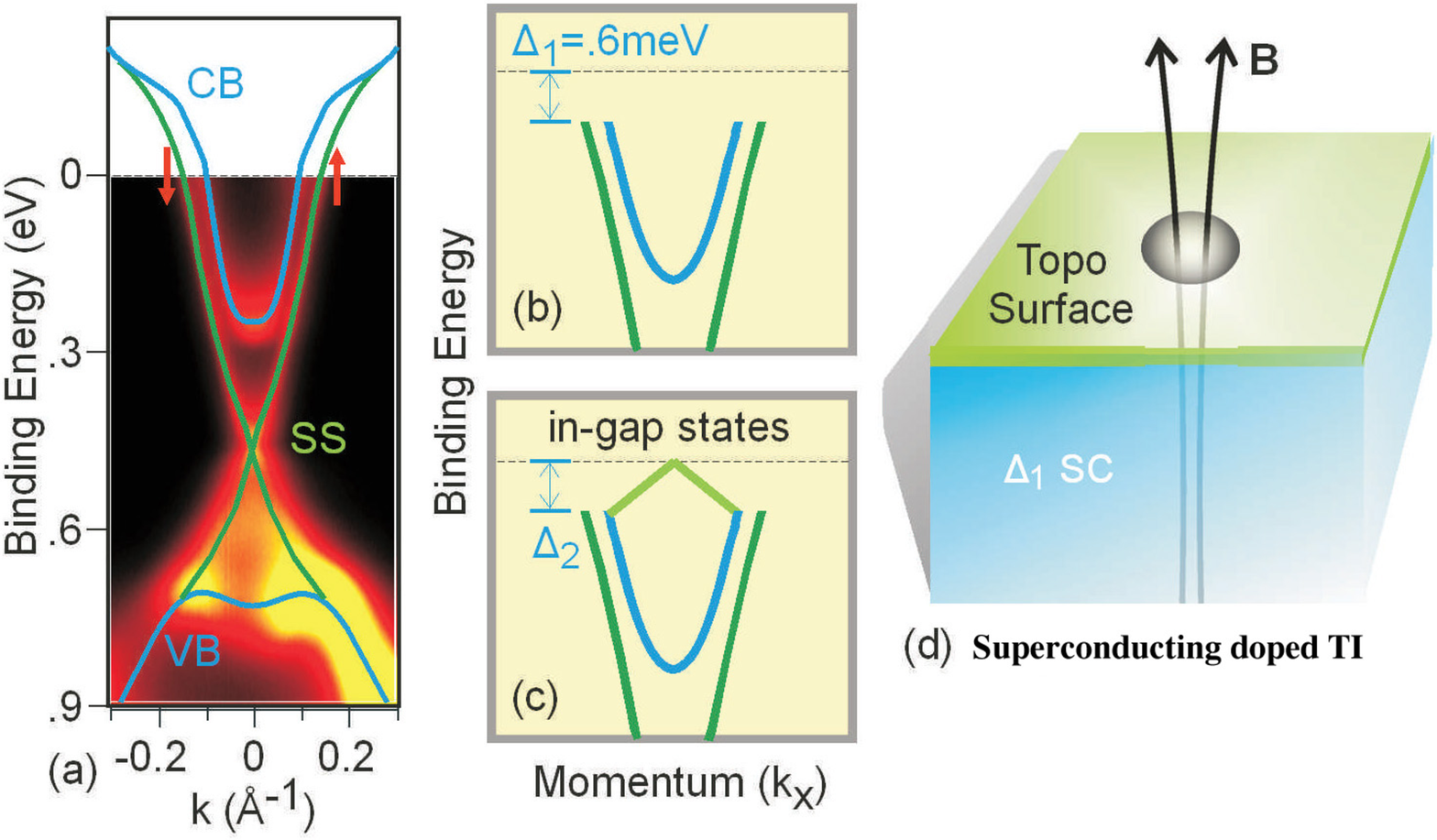}
\caption{\label{MF} \textbf{Superconducting doped topological insulator.} (a) Topologically protected surface states cross the Fermi level before merging with the bulk valence and conduction bands in a lightly doped topological insulator. (b) If the superconducting wavefunction has even parity, the surface states will be gapped by the proximity effect, and vortices on the crystal surface will host braidable Majorana fermions. (c) If superconducting parity is odd, the material will be a so-called "topological superconductor", and new states will appear below T$_c$ to span the bulk superconducting gap. (d) Majorana fermion surface vortices are found at the end of bulk vortex lines and could be manipulated for quantum computation if superconducting pairing is even. [Adapted from L. Wray $et$ $al$., \textit{Nature Phys.} \textbf{6}, 855 (2010). \cite{Wray1}]} \end{figure*}

The interplay between the topological order and superconductivity may lead to many proposals of novel quantum phenomena such as time-reversal invariant topological superconductors \cite{Kane_Proximity, Zhang_TSC, Fu_CuBiSe}, Majorana fermions
\cite{Kane_Proximity, Zhang_TSC, Fu_CuBiSe}, and fault-tolerant quantum computation \cite{Zhang_TSC, Fu_CuBiSe}. Currently, researchers have been focused on two approaches to introduce superconductivity into the a TI. The first approach is to bulk dope a TI material in order to make it a bulk superconductor. The most notable example is the bulk superconductivity with $T_{\textrm{c}}\sim3.8$ K found in copper-doped bismuth selenide Cu$_{0.12}$Bi$_2$Se$_3$. The second approach is to utilize the superconducting proximity effect by interfacing a TI with a superconductor. We review the ARPES studies for both approaches as follows.

The bulk superconductivity reported in copper-doped bismuth selenide Cu$_{x}$Bi$_2$Se$_3$ \cite{Hor} has attracted much interests \cite{Hor, Wray1, Ando_CuBiSe, Shin_CuBiSe, CuBiSe_STM}. A major contribution made by ARPES measurements \cite{Wray1} is that ARPES shows that the topological surface states remain well defined and non-degenerate with bulk electronic states at the Fermi level of optimally doped superconducting Cu$_{0.12}$Bi$_2$Se$_3$. This observation is important for the following reasons: Since the bulk is superconducting, then it is possible to use the natural proximity effect between the bulk and surface to induce superconductivity on the surface of Cu$_{0.12}$Bi$_2$Se$_3$. The superconductivity in these spin-helical Dirac surface states can realize a 2D topological superconductor. And Majorana fermion bound states may exist in the magnetic vortices at the surface schematically shown in Fig.\ref{MF}. However, this exciting scenario is only possible if the surface states are non-degenerate with the bulk bands at the Fermi level. Because otherwise the surface and bulk superconductivity are strongly coupled and the Majorana fermion bound state trapped in a surface vortex can leak to the bulk, causing decoherence and annihilation of the Majorana fermion. Therefore, the ARPES observation of non-degenerate nature of the surface states at the Fermi level of optimally doped superconducting Cu$_{0.12}$Bi$_2$Se$_3$ reported in \cite{Wray1} serves as the key for realizing topological superconductivity on the surface of superconducting Cu$_{0.12}$Bi$_2$Se$_3$.

The superconductivity physics in Cu$_{x}$Bi$_2$Se$_3$ can be even richer. In Ref. \cite{Fu_CuBiSe}, the authors proposed the theoretical possibility that the bulk superconductivity in Cu$_{x}$Bi$_2$Se$_3$ may also be topologically nontrivial. If the intra- and inter-orbital hopping parameters lie in an appropriate regime, theory in Ref. \cite{Fu_CuBiSe} shows that the Cu$_{x}$Bi$_2$Se$_3$ system is a bulk odd-parity topological superconductor, and one would expect helical Majorana surface states. However, the nature of the bulk superconductivity is still controversial. Although zero-bias peak in a point contact experiment has been reported and interpreted as the signature for the bulk topological superconductivity, high-resolution ARPES \cite{Shin_CuBiSe} in fact did not resolve any observable superconducting gap neither in the bulk bands nor in the surface states, and STM measurements \cite{CuBiSe_STM} suggest the pairing Cu$_{x}$Bi$_2$Se$_3$ seems to be conventional (topologically trivial). Furthermore, recent theoretical and experimental studies suggest that the nature of the zero-bias peak can be very complex \cite{Patrick Lee2, TeWari, Marcus, Franceschi}, which therefore cannot serve as conclusive signature for the topological superconductivity or the Majorana fermions.

As for the superconducting proximity effect approach, there have been many transport and STM studies on this topic \cite{Nitin, Kanigel, Morpurgo, LuLi1, Mason, LuLi2, Gordon, Dong, Brinkman, Burch, Molenkamp, Kirzhner, Millo, LuLi3}. However, due to the lack of momentum and spin-resolution of transport and STM, these experiments cannot show that the topological surface states are indeed superconducting since topological surface states, bulk bands, and potentially trivial surface states or impurity states all contribute to the transport or STM signals. ARPES studies on TI/superconductor proximity effect samples are on the other hand very limited and under debate \cite{BSCCO_Hasan, BSCCO_Valla}. Although extensive experimental efforts are underway, critical signatures regarding the observation of unambiguous Majorana mode are still lacking. Without the demonstration of helical Cooper pairing in the topological (spin only) Dirac surface states (which can presently be done only via ARPES thanks to its momentum and \textit{spin}-resolution and buried interface sensitivity), critical evidence for time-reversal invariant topological superconductivity (TRI-TSC) is still lacking.

\bigskip
\bigskip

\textbf{Acknowledgement}
The authors acknowledge N. Alidoust, A. Bansil, I. Belopolski, B. A. Bernevig, G. Bian,  R. J. Cava, F. C. Chou, J. H. Dil, A. V. Fedorov, C. Fang, Liang Fu, S. Jia, C. L. Kane, D. Hsieh, Y. S. Hor, H. Lin, C. Liu, D. Qian, J. Osterwalder, A. Richardella, N. Samarth, R. Sankar, A. Vishwanath, L. A. Wray, and Y. Xia for collaboration and U.S. DOE DE-FG-02-05ER46200, No. AC03-76SF00098, and No. DE-FG02-07ER46352 for support. M.Z.H. acknowledges visiting-scientist support from Lawrence Berkeley National Laboratory and additional support from the A.P. Sloan Foundation and Princeton University.

\end{document}